\documentclass[11pt,a4paper]{article}
\usepackage{jcappub}
\usepackage[T1]{fontenc}
\usepackage{xcolor}
\usepackage{graphicx}
\usepackage{bm}
\usepackage{amsmath,amsfonts,amssymb}
\usepackage{mathtools}
\usepackage{braket}
\usepackage[normalem]{ulem}
\usepackage{setspace} 
\usepackage{here}
 
\begin{document}


\title{Revisiting the Affleck-Dine mechanism for primordial black hole formation}
\author[a]{Kentaro Kasai,}
\author[a,b]{Masahiro Kawasaki,}
\author[a,b]{and Kai Murai}
\affiliation[a]{ICRR, University of Tokyo, Kashiwa, 277-8582, Japan}
\affiliation[b]{Kavli IPMU (WPI), UTIAS, University of Tokyo, Kashiwa, 277-8583, Japan}

\abstract{%
We study a primordial black hole (PBH) formation scenario based on the Affleck-Dine (AD) mechanism and investigate two PBH mass regions: $M \sim 30 M_\odot$ motivated by the LIGO-Virgo observations of the binary black hole mergers and $M \gtrsim 10^4 M_\odot$ motivated by the observations of supermassive black holes at the center of galaxies.
In the previous studies, it has been considered that the inhomogeneous AD baryogenesis generates regions with a large baryon asymmetry, some of which collapse into PBHs.
In this paper, we show that this scenario is severely constrained due to the baryon asymmetry remaining outside PBHs, which would spoil the success of the big bang nucleosynthesis.
Then, we propose an alternative scenario where the AD leptogenesis results in the inhomogeneous formation of Q-balls with lepton charges, which collapse into PBHs.
As a result, we find that our scenario can explain the favorable PBH abundance without conflicting with the observational constraints.
}

\emailAdd{kkasai@icrr.u-tokyo.ac.jp}
\emailAdd{kawasaki@icrr.u-tokyo.ac.jp}
\emailAdd{kmurai@icrr.u-tokyo.ac.jp}

\maketitle

\section{Introduction}
\label{sec: intro}

Recent LIGO-Virgo observations have revealed more than 50 binary mergers of black holes (BHs) with $M\sim30M_{\odot}$~\cite{LIGOScientific:2016aoc,LIGOScientific:2018mvr,LIGOScientific:2020ibl,LIGOScientific:2021usb,LIGOScientific:2021djp}, which has been of great interest in astrophysics and cosmology.
In addition, supermassive black holes (SMBHs), which have masses larger than $10^5M_{\odot}$, are thought to exist at the centers of almost all galaxies~\cite{Kormendy:1995er,Magorrian:1997hw,Richstone:1998ky}.
In particular, SMBHs of $M\gtrsim10^9 M_{\odot}$ have been found at $z \sim 6$ (e.g.,~\cite{Matsuoka:2016pho,Banados:2017unc}), which severely constrains the physical mechanism of their formation~\cite{Volonteri:2010wz}.
These black holes can have a primordial origin~\cite{Bird:2016dcv,Kashlinsky:2016sdv,Sasaki:2016jop,Carr:2016drx,Clesse:2016vqa,Eroshenko:2016hmn,Ali-Haimoud:2017rtz,Liu:2018ess,Vaskonen:2019jpv,Liu:2019rnx,Wu:2020drm,DeLuca:2020bjf,DeLuca:2020qqa}, and are called primordial black holes (PBHs)~\cite{Hawking:1971ei,Carr:1974nx,Carr:1975qj}.
PBHs are formed by the gravitational collapse of large density fluctuations on small scales during the radiation-dominated era.

In order to produce a significant number of PBHs, the amplitude of the density fluctuations should be much larger than that observed on the CMB scale, which is difficult to produce in a single-field inflation model.
The inflation models producing such large density fluctuations were studied in Refs.~\cite{Garcia-Bellido:1996mdl,Yokoyama:1995ex,Kawasaki:1997ju}, subsequently in Refs.~\cite{Kawasaki:2016pql,Inomata:2016rbd,Inomata:2018cht} (for a review, see Ref.~\cite{Sasaki:2018dmp}). 
There were also PBH formation models where the curvaton or spectator field produces large density fluctuations~\cite{Ando:2017veq,Ando:2018nge,Kohri:2012yw, Chen:2019zza, Liu:2021rgq, Pi:2021dft, Suyama:2011pu, Maeso:2021xvl, Cai:2021wzd, Carr:2017edp}.

However, the PBH formation from the initial curvature fluctuations is indirectly subject to several observational constraints~\cite{Carr:2020gox}.
The most stringent constraint in the PBH mass range of $10^4M_{\odot}\lesssim M \lesssim 4\times10^{13}M_{\odot}$ is the $\mu$-distortion of the CMB spectrum due to the Silk damping of small-scale density perturbations~\cite{Chluba:2012we,Kohri:2014lza}.
This excludes curvature fluctuations that generate PBHs with the mass range mentioned above.
Furthermore, the small-scale density fluctuations that form PBHs produce background gravitational waves (GWs) through second-order effects of the curvature perturbations, which dominate over the primordial GWs produced during inflation~\cite{Saito:2008jc,Bugaev:2010bb}.
The null detection of those gravitational waves by the pulsar timing array experiments~\cite{Lentati:2015qwp,NANOGrav:2015aud,Shannon:2015ect} also excludes curvature fluctuations generating PBHs of $0.1M_{\odot} \lesssim M\lesssim 10M_{\odot}$.
These are different from direct constraints on PBHs such as Refs.~\cite{Lu:2020bmd,Oguri:2017ock,Ali-Haimoud:2016mbv,Murgia:2019duy,Brandt:2016aco,Koushiappas:2017chw,Monroy-Rodriguez:2014ula,Smith:2016hsc,Carr:2018rid,Ricotti:2007au}.

These constraints can be weakened or evaded by using density fluctuations with non-Gaussianity~\cite{Garcia-Bellido:2017aan} or nonlinear local objects~\cite{Hawking:1987bn,Caldwell:1995fu,Garriga:1992nm,Cotner:2016cvr,Nakama:2016kfq,Deng:2017uwc,Kitajima:2020kig,Kawana:2021tde,Huang:2022him,Cotner:2018vug,Cotner:2019ykd,Kusenko:2020pcg}.
One of the latter mechanisms is the PBH formation model using Affleck-Dine (AD) baryogenesis~\cite{Dolgov:1992pu,Dolgov:2008wu,Hasegawa:2017jtk,Hasegawa:2018yuy,Kawasaki:2019iis}
The idea that inhomogeneous baryogenesis produces large density fluctuations to lead PBHs was first proposed in Ref.~\cite{Dolgov:1992pu} and was further studied in Ref.~\cite{Dolgov:2008wu}.
However, the model was ad hoc, and the PBH mass function was assumed to be a log-normal form, whose relation to the model parameters was unclear.  
The inhomogeneous AD baryogenesis model based on the framework of supersymmetry (SUSY) was proposed in Refs.~\cite{Hasegawa:2017jtk,Hasegawa:2018yuy}, where the PBH mass function is calculable for a given set of model parameters. 
The model was applied to the LIGO-Virgo GW events~\cite{Hasegawa:2017jtk,Hasegawa:2018yuy} and was then used to explain supermassive black holes at the center of galaxies~\cite{Kawasaki:2019iis}. 

In AD baryogenesis model, one of the flat directions (called AD field) in the minimum supersymmetric standard model (MSSM) is considered, and the dynamics of the AD field leads to the PBH formation.
The coarse-grained AD field exhibits stochastic behavior, whose evolution is described by the Fokker-Planck equation.
In regions where the averaged AD field exceeds a certain threshold value, a large baryon asymmetry is generated through the subsequent evolution of the AD field.
Those regions with large baryon number are called high baryon bubbles (HBBs). 
On the other hand, in regions with the AD field less than the threshold value, the AD field rolls down to the origin and produces no baryon asymmetry.
The HBBs then gravitationally collapse into PBHs.
Furthermore, this model predicts strong clustering of PBHs, which produces significant isocurvature fluctuations and changes the merger rate~\cite{Kawasaki:2021zir}.

However, in this model, small HBBs do not gravitationally collapse and remain as regions with very high baryon density.
This leads to a highly inhomogeneous baryon distribution with average baryon density much larger than the observed one, which would spoil the success of the big bang nucleosynthesis (BBN).
This difficulty could be avoided if there exist HBBs with both positive and negative baryon charges, and nucleons and anti-nucleons diffusing from HBBs annihilate each other.
However, as we show in this paper, this does not work because regions with positive baryon charges are far apart from those with negative baryon charges compared to the typical diffusion length of nucleons.

Therefore, this paper proposes an alternative modification of the AD mechanism for the PBH formation.
The model utilizes a flat direction with lepton number (such as $LL\bar{e}$) instead of a baryonic flat direction, and produces high lepton bubbles (HLB). 
In addition, we consider the formation of L-balls~\cite{Kawasaki:2002hq} (Q-balls~\cite{Coleman:1985ki,Kusenko:1997zq,Kusenko:1997si,Enqvist:1997si,Kasuya:1999wu} with lepton charges) in the HLBs.
A large number of L-balls aggregate spherically in each HLB and then gravitationally collapse into PBHs.
The lepton charge in the L-balls is protected from the sphaleron processes by which lepton numbers are partially converted to baryon numbers.
Thus, this model can avoid the inhomogeneous baryon density problem.
However, some fraction of the lepton charge evaporates from L-balls~\cite{Laine:1998rg} and it produces a baryon number through the sphaleron processes. 
We quantitatively estimate the produced baryon number density and examine whether it is consistent with the observation.
The L-balls eventually decay by emitting neutrinos~\cite{Cohen:1986ct,Kawasaki:2012gk}.
We also study the injected non-thermal neutrinos and its effect on the effective number of neutrino species.

This paper is organized as follows.
First, in Sec.~\ref{sec2}, we review the scenario of PBH formation based on the AD baryogenesis.
Next, in Sec.~\ref{sec3}, we show that the baryon number generated from each HBB remains without pair annihilation, which leads to inconsistency with BBN.
In Sec.~\ref{sec4}, we present an alternative scenario with L-balls.
Sec.~\ref{sec5} is devoted to the conclusion.

\section{Primordial Black Holes from Affleck-Dine mechanism}
\label{sec2}

In this section, we briefly review the Affleck-Dine mechanism for the PBH formation in the MSSM~\cite{Hasegawa:2017jtk,Hasegawa:2018yuy,Kawasaki:2019iis,Kawasaki:2021zir}. 

\subsection{Affleck-Dine mechanism}

We consider the model based on the AD mechanism~\cite{Affleck:1984fy,Dine:1995kz}, which utilizes one of the flat directions (an AD field) in the MSSM.
The potential of the AD field is lifted by SUSY breaking effects and the existence of a cutoff ($=$Planck scale $M_\text{Pl}$).
During inflation, the potential is given by
\begin{align}
    V(\phi) 
    & = m_\phi^2|\phi|^2 - cH^2|\phi|^2
    + V_\text{NR} + V_\text{A}\nonumber\\
    & = m_\phi^2|\phi|^2 - cH^2|\phi|^2
    + |\lambda|^2 \frac{|\phi|^{2(n-1)}}{M_{\rm{Pl}}^{2(n-3)}} 
    + \left(\lambda a_M \frac{m_{3/2}\phi^n}{n M_{\rm{Pl}}^{n-3}}
    + \mathrm{h.c.}\right)
    ,
\end{align}
where $m_\phi$ is a soft SUSY breaking mass, $H$ is the Hubble parameter, $m_{3/2}$ is the gravitino mass, $c$, $a_\text{M}$, and $\lambda$ are dimensionless constants.
The first and second terms represent the soft SUSY breaking mass term and the Hubble induced mass term, respectively.
The third term $V_\text{NR}$ is called non-renormalizable term, and the fourth term $V_\text{A}$ is the A-term, which violates a global $U(1)$ symmetry.
When the Hubble induced mass term is negative ($c > 0$), the AD field is destabilized at the origin and has a large field value during inflation.
After inflation, the soft SUSY breaking term overcomes the Hubble induced mass term, and the AD field starts to oscillate.
At that time, the AD field is kicked in the phase direction $\theta$ due to the A-term, and hence the AD field rotates in the complex field space, which leads to baryogenesis or leptogenesis since the angular momentum $\sim  |\phi|^2\dot{\theta}$ corresponds to a baryon or lepton number density.
This is the conventional AD baryogenesis (leptogenesis).

In the AD mechanism for the PBH formation, we make the following two unconventional assumptions:
\begin{itemize}
    \item[(i)]
    During inflation, the AD field has a positive Hubble-induced mass while it has a negative one after inflation.
    \item[(ii)] 
    After inflation, the thermal potential for the AD field overcomes the negative Hubble-induced mass around the origin of the field space.
\end{itemize}
These assumptions can be satisfied by choosing the couplings of the AD field to the potential and kinetic terms of the inflaton.
The scalar potential for the AD field $\phi \equiv \varphi e^{i\theta}$ is given by
\begin{equation}
    V(\phi)
    =
    \left\{
        \begin{array}{ll}
        (m_\phi^2+c_I H^2)|\phi|^2+V_{\rm{NR}} +V_\text{A}
        & \quad
        ({\rm{During\ inflation}})
        \\
        (m_\phi^2-c_M H^2)|\phi|^2+V_{\rm{NR}}+V_\text{A}+V_{\rm{T}}(\phi)
        & \quad
        ({\rm{After\ inflation}})
        \end{array}
    \right.
    ,
    \label{potential}
\end{equation}
where $c_I$ and $c_M$ are positive and dimensionless constants.
$V_{\rm{T}}$ is the thermal potential for the AD field due to interactions with decay products of the inflaton and is written as
\begin{equation}
    V_{\rm{T}}(\phi)= \left\{
    \begin{array}{ll}
        c_1 T^2|\phi|^2
        & \quad
        (|\phi| \lesssim T)
        \\
        c_2 T^4\ln \left(\frac{|\phi|^2}{T^2}\right)
        & \quad
        (|\phi| \gtrsim T)
    \end{array}
    \right.
    ,
    \label{thermal potential}
\end{equation}
where $c_1$ and $c_2$ are $\mathcal{O}(1)$ parameters.

When the above two conditions are satisfied, the AD potential~\eqref{potential} has an interesting feature.
During inflation, the potential has a unique minimum at the origin due to the positive Hubble-induced mass.
On the other hand, after inflation, the potential has two local minima at $\phi=0$ and $|\phi|\neq 0$ for the following reasons.
Due to condition (ii), the origin becomes a local minimum.
For $|\phi| \gtrsim T$, the thermal potential becomes subdominant compared to the Hubble-induced mass.
Then, the balance between the Hubble-induced mass and the non-renormalizable term induces the other local minimum.
The field value at the latter local minimum is given by
\begin{align}
    |\phi| 
    \simeq
    \varphi_m
    \equiv
    \left( \frac{HM_{\rm{Pl}}^{n-3}}{|\lambda|} \right)^{\frac{1}{n-2}}
    .
\end{align}
The local maximum between the origin and $|\phi|\sim \varphi_m$ is determined by the balance between the Hubble-induced mass and the thermal potential as 
\begin{align}
    |\phi|
    \simeq
    \varphi_c
    \equiv
    \sqrt{\frac{c_2}{c_M}} \frac{T(t)^2}{H(t)}
    ,
\end{align}
where we assume $V(\varphi)\simeq V_{\rm{T}}(\varphi)-c_MH^2\varphi^2\simeq c_2T^4-c_MH^2\varphi^2$ around $\varphi\sim\varphi_c$.
To satisfy condition (ii), $T(t)>H(t)$ must always be satisfied after inflation.
This condition can be read as 
\begin{align}
    \Delta
    \equiv
    \frac{T_{\rm{RH}}^2 M_{\rm{Pl}}}{H(t_e)^3}
    \gtrsim 1
    ,
\end{align}
where $T_{\rm{RH}}$ is the reheating temperature and $t_e$ is the time at the end of inflation.
Here we assume that the evolution of the temperature just after inflation is given by
\begin{equation}
    T(t)
    \sim
    \left( T_{\rm{RH}}^2H(t)M_{\rm{Pl}} \right)^{1/4}
    .
\end{equation}

\subsection{Formation of high baryon bubbles}
\label{Formation of high baryon bubbles}

In the remainder of this section, we follow the previous studies~\cite{Hasegawa:2017jtk,Hasegawa:2018yuy,Kawasaki:2019iis,Kawasaki:2021zir} and focus on the model with the Affleck-Dine baryogenesis, not the leptogenesis.
Immediately after inflation, the AD field has a multiple vacuum structure.
We denote the origin by ``vacuum A'' and the minimum at $\varphi \simeq \varphi_m$ by ``vacuum B''.
\begin{figure}[htbp]
    \centering
    \includegraphics[width=0.72\linewidth]{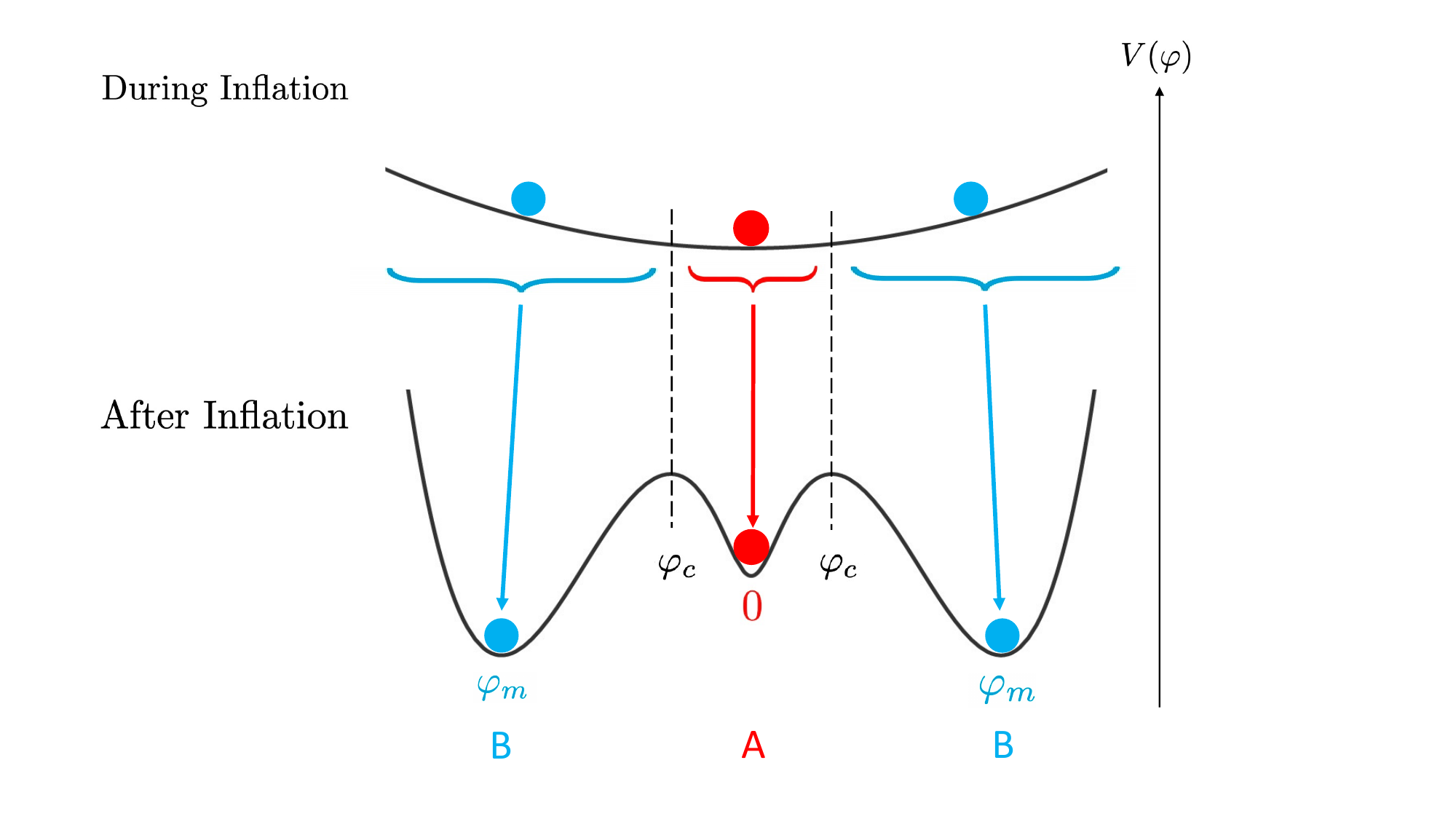}
    \caption{
        Schematic view of the dynamics of the AD field during and after inflation.
        \textit{Upper side}: During inflation, the IR modes of the AD field diffuse in the complex plane and take different values in different Hubble patches.
        \textit{Lower side}: Just after inflation, the multi-vacuum structure is realized. 
        In the regions with $|\phi|\lesssim \varphi_c$, $\phi$ rolls down to vacuum A and almost no baryon asymmetry is generated.
        On the other hand, if $|\phi|\gtrsim \varphi_c$, $\phi$ classically rolls down to vacuum B and are identified as HBBs later.
        }
    \label{fig:AD field potential}
\end{figure}

Due to the quantum fluctuations, $\phi$ takes different values in different places at the end of inflation.
If $|\phi|$ is smaller than $\varphi_c$ at the end of inflation, the AD field rolls down to vacuum A and no baryon number is generated.

On the other hand, in the region where $|\phi|$ exceeds $\varphi_c$, the AD field rolls down to vacuum B after inflation.
After a while, the SUSY breaking term overcomes the negative Hubble-induced term, and vacuum B becomes unstable.
Then, the AD field starts to oscillate around the origin and creates large baryon number due to the phase-dependent A-term as in the conventional AD mechanism. 
In this way, high baryon bubbles (HBBs) are produced.

The baryon-to-entropy ratio generated in each HBB is given by 
\begin{equation}
    \eta_b
    \equiv
    \frac{n_b}{s}
    \simeq
    \epsilon \frac{T_{\rm{RH}}m_{3/2}}{H_{\rm{osc}}^2}
    \left( \frac{\varphi_{\rm{osc}}}{M_{\rm{Pl}}} \right)^2
    ,
    \label{baryon number made from AD mechanism}
\end{equation}
where $n_b$ is the baryon number density, $s$ is the entropy density, and $\varphi_{\rm{osc}}$ is the field value when the AD field starts to oscillate. 
$\epsilon$ represents an efficiency of the asymmetry generation and is given by
\begin{equation}
    \epsilon
    =
    \sqrt{\frac{c_M}{n-1}}
    \frac{ q_b |a_M| \sin[n\theta_0 + \arg(a_M)] }{ 3\left( \frac{n-4}{n-2}+1 \right) }
    ,
    \label{value_of_epsilon}
\end{equation}
where $q_b$ is the baryon charge of the AD field, and $\theta_0$ is the initial phase of the AD field.

\subsection{PBH formation}

Next, we briefly introduce the mechanism of PBH formation and show the dependence of the PBH abundance on the parameter $c_I$ in advance.
As we saw in Sec.~\ref{Formation of high baryon bubbles}, HBBs are formed in the region where the AD field exceeds $\varphi_c$ at the end of inflation.

The HBBs that we are interested in have superhorizon sizes when they are formed.
The density contrast between inside and outside the HBB increases after non-relativistic nucleons or Q-balls form in the HBBs.
As shown later, when the HBBs enter the horizon, they collapse into PBHs if their density contrasts exceed a certain threshold value.
On the other hand, small HBBs with density contrasts less than the threshold do not form PBHs.

If the Hubble-induced mass potential is steep during inflation, the AD field is localized around the origin in the complex field space and the probability that the AD field exceeds $\varphi_c$ is suppressed.
Therefore, in order to obtain the sufficient abundance of PBHs, we favor $c_I\ll 1$.

\subsubsection{The case with nucleon formation}
\label{Without stable Q-ball formation}

At a high temperature, the baryon number in HBBs is carried by quarks, and they remain relativistic.
Thus, the energy density inside HBBs is the same as that outside HBBs.
After the QCD phase transition, the baryon number is carried by non-relativistic nucleons.
Then, the density contrast between inside and outside HBBs grows and is estimated by
\begin{align}
    \delta
    \equiv
    \frac{ \rho^{\rm{in}}-\rho^{\rm{out}} }{ \rho^{\rm{out}} }
    &\simeq
    \frac{ n_b m_b }{ \pi^2 g_{\ast} T^4 / 30 } \theta(T_{\rm{QCD}}-T)
\nonumber\\
    &\simeq
    6.3\eta_b\left( \frac{T}{200~{\rm{MeV}}} \right)^{-1} 
    \theta(T_{\rm{QCD}}-T)
    \label{eq : density contrast in nucleon bubble model}
    ,
\end{align}
where $m_b \simeq 938~{\rm{MeV}}$ is the nucleon mass and $T_\mathrm{QCD}$ is the temperature at the QCD phase transition.

We regard this density contrast as measured in a comoving slice.
Then, the threshold value of the density contrast for the PBH formation is given by~\cite{Harada:2013epa}
\begin{equation}
    \delta_c
    \simeq
    \frac{3(1+w)}{5+3w}
    \sin^2 \left( \frac{\pi\sqrt{w}}{1+3w} \right)
    ,
\end{equation}
where $w$ is the parameter of the equation of state in HBBs and written as
\begin{align}
    w
    =
    \frac{p^{\rm{in}}}{\rho^{\rm{in}}}
    \simeq
    \frac{p^{\rm{out}}}{\rho^{\rm{in}}}
    =
    \frac{1}{3(1+\delta)}
    .
\end{align}
Therefore, we obtain the condition for the PBH formation as
\begin{equation}
    \delta
    \gtrsim
    \frac{4+3\delta}{6+5\delta}\sin^2 \left( \sqrt{\frac{1+\delta}{3}} \frac{\pi}{2+\delta} \right)
    \iff 
    \delta
    \gtrsim 0.40
    .
    \label{condition for PBH formation}
\end{equation}
This condition gives the upper bound of the temperature at the horizon reentry for the PBH formation as
\begin{align}
    T_c
    \simeq
    {\rm{min}}[3.2\eta_b~{\rm{GeV}},\,T_{\rm{QCD}}]
    .
    \label{Threshold temperature}
\end{align}
Now, we comment on the relations among the PBH mass $M$, the temperature $T$ at the PBH formation, the wave number $k$ of the corresponding fluctuation mode, and the e-fold number $N_k$ when this mode exits the horizon during inflation.
The PBH mass is roughly given by the horizon mass when the fluctuations with $k$ reenter the horizon and hence written as 
\begin{equation}
   M(k)
   \simeq
   20.5M_{\odot} \left(\frac{g_{\ast}}{10.75} \right)^{-1/6}
   \left( \frac{k}{10^6~{\rm{Mpc}}^{-1}} \right)^{-2},
   \label{Hubble mass with respect to k}
\end{equation}
where $g_{\ast}$ is the effective degrees of freedom of relativistic particles. The PBH formation temperature is given by 
\begin{equation}
   T(k)
   \simeq
   85.5~{\rm{MeV}}
   \left( \frac{g_{\ast}}{10.75} \right)^{-1/6}
   \left( \frac{k}{10^6~{\rm{Mpc}}^{-1}} \right).
   \label{temperature with respect to k}
\end{equation}
Moreover, the e-folding number $N_k$ at the horizon crossing during inflation is written as
\begin{equation}
    N_k=\ln \left( \frac{k}{k_0} \right),
    \label{e-fold number with respect to k}
\end{equation}
where $k_0=2.24\times10^{-4}~{\rm{Mpc}^{-1}}$.
Using Eqs.~\eqref{Threshold temperature}, \eqref{Hubble mass with respect to k}, and \eqref{temperature with respect to k}, we obtain the lower bound for the PBH mass as
\begin{align}
    M
    \gtrsim
    M_c
    \equiv
    {\rm{max}}
    \left[
        1.5\times10^{-2} \eta_b^{-2},\,
        3.75 \left( \frac{T_{\rm{QCD}}}{200~{\rm{MeV}}} \right)^{-2}
    \right]
    \times \left( \frac{g_{\ast}}{10.75} \right)^{-1/2}M_{\odot}
    .
\end{align}
In other words, regions whose horizon mass is smaller than this threshold mass do not collapse into PBHs and remain as HBBs after the horizon reentry.

\subsubsection{Case with stable Q-ball formation}

In the above discussion, we assumed that the AD field directly decays into SM particles.
However, there are cases where the AD field has a spatial instability and forms non-topological solitons called Q-balls just after the onset of oscillations.
A Q-ball is a spherically symmetric field configuration that minimizes the energy with a fixed $U(1)$ charge.
The charge and energy inside a Q-ball depend on the SUSY breaking schemes. 

In the gravity-mediated SUSY breaking scenario, Q-balls with baryon charge are unstable and decay into quarks and lightest SUSY particles (LSPs) before the QCD phase transition.
Since the Q-balls are non-relativistic, the density contrast between inside and outside of the HBBs increases until the Q-ball decay, and it grows again when the LSPs become non-relativistic.
Thus, the situation is much more complicated than the case without Q-ball formation.
Furthermore, the produced LSPs may give too much contribution to the present matter density.
Therefore, we do not consider this scenario here. 

On the other hand, Q-balls with baryon charge are stable in the gauge-mediated SUSY breaking scenario.
In this case, we need to evaluate the evolution of Q-balls.
We represent the abundance of stable Q-balls formed inside the HBBs as 
\begin{equation}
    Y_Q^{\rm{in}}
    \equiv
    \frac{\rho_Q^{\rm{in}}}{s}
    =
    \omega_Q\eta_b^{\rm{in}}
    ,
    \label{definition of Y_Q}
\end{equation}
where $\rho_Q^\mathrm{in}$ is the energy density of Q-balls inside the HBBs, $\omega_Q$ is the energy per unit $U(1)$ charge inside the Q-ball and $\eta_b^{\rm{in}}$ is the charge density inside the HBBs. 
In the gauge-mediated SUSY breaking, $\omega_Q$ is given by the gravitino mass $m_{3/2}$ for Q-balls with large charges.
Then, the density contrast becomes
\begin{equation}
    \delta
    =
    \frac{\rho_Q^{\rm{in}}}{(\pi^2/30)g_{\ast}T^4}
    =
    \frac{4}{3}\left( \frac{T}{Y_Q^{\rm{in}}} \right)^{-1}
    .
    \label{density contrast}
\end{equation}
With a similar calculation to Sec.~\ref{Without stable Q-ball formation}, this leads to the threshold mass $M_c$ given by
\begin{equation}
    M_c
    =
    3.75M_{\odot}
    \left( \frac{g_{\ast}}{10.75} \right)^{-1/4}
    \left( \frac{Y_Q^{\rm{in}}}{60~{\rm{MeV}}} \right)^{-2}
    .
    \label{threshold mass}
\end{equation}
Importantly, the discussion to obtain Eqs.~\eqref{definition of Y_Q}, \eqref{density contrast}, and \eqref{threshold mass} is also applied to the Affleck-Dine leptogenesis if we replace the baryon charge with the lepton one.

\subsection{Distribution of HBBs and Mass spectrum of PBH}
\label{Distribution of HBBs}

To estimate the size distribution of HBBs, we consider the stochastic behavior of the IR modes of the AD field during inflation.
The probability distribution function of the coarse-grained field value $\phi$ at the e-folding number $N$ obeys the Fokker-Planck equation~\cite{Starobinsky:1982ee,Linde:1982uu,Starobinsky:1994bd} given by
\begin{equation}
    \frac{\partial P(N,\phi)}{\partial N}
    =
    \sum_{i=1,2} \frac{\partial}{\partial\phi_i}
    \left[ 
        \frac{\partial_{\phi_i }V P(N,\phi)}{3H_I^2}
        + \frac{H_I^2}{8\pi^2}\frac{\partial P(N,\phi)}{\partial\phi_i}
    \right]
    ,
    \label{Fokker Planck equation}
\end{equation}
where $(\phi_1,\phi_2)=\sqrt{2}\,({\rm{Re}}[\phi],{\rm{Im}}[\phi])$.
The first term in the RHS represents the classical force induced by the potential $V(\phi)$ and the second term represents the Gaussian noise due to the quantum fluctuations.
Since we are interested in the time evolution of $\phi$ during inflation, we approximate the potential by
\begin{align}
    V(\phi)
    =
    \frac{1}{2}c_I H_I^2 \left( \phi_1^2 + \phi_2^2 \right)
    .
\end{align}
Here, we ignore the contribution of $V_{\rm{NR}}$, $V_\text{A}$, and the soft SUSY breaking mass term.
In the calculation below, we assume that the Hubble parameter is constant during inflation.
Let us rewrite the equation using $(\varphi,\theta) \equiv (|\phi|,{\rm{arg}}\phi)$.
In addition, we define the dimensionless field $\tilde{\varphi}\equiv2\pi\varphi/H_I$ and define $\tilde{P}(N,\tilde{\varphi})$ by integrating over the phase direction as 
\begin{align}
    \tilde{P}(N,\tilde{\varphi})
    \equiv
    \left( \frac{H_I}{2\pi} \right)^2
    \tilde{\varphi}\int_{0}^{2\pi} \mathrm{d}\theta \, P(N,\varphi,\theta)
    .
\end{align}
Then, $\tilde{P}(N,\tilde{\varphi})$ satisfies
\begin{align}
    \frac{\partial\tilde{P}(N,\tilde{\varphi})}{\partial N}
    =
    \frac{c_I'}{2}\left(
        \tilde{P}(N,\tilde{\varphi})
        + \tilde{\varphi} \frac{\tilde{P}(N,\tilde{\varphi})}{\partial \tilde{\varphi}}
    \right)
    +\frac{1}{4}\left(
        \frac{\partial^2\tilde{P}(N,\tilde{\varphi})}{\partial\tilde{\varphi}^2}
        - \frac{1}{\tilde{\varphi}}
          \frac{\partial \tilde{P}(N,\tilde{\varphi})}{\partial \tilde{\varphi}}
        + \frac{\tilde{P}(N,\tilde{\varphi})}{\tilde{\varphi}^2}
    \right)
    ,
\end{align}
where $c_I' \equiv 2c_I/3$. 
For the initial condition $\tilde{P}(N=0,\tilde{\varphi})=\delta(\tilde{\varphi})$, the solution is written as
\begin{equation}
    \tilde{P}(N,\tilde{\varphi})
    =
    \frac{\tilde{\varphi}}{\tilde{\sigma}^2(N)}
    \exp\left[ -\frac{\tilde{\varphi}^2}{2\tilde{\sigma}^2(N)} \right]
    ,
\end{equation}
where $\tilde{\sigma}^2(N) \equiv (1-e^{-c_I'N})/2c_I'$.

Let us focus on a fixed comoving patch that exits the horizon at the e-folding number $N$.
At the horizon exit, the probability distribution of $\tilde{\varphi}$ averaged over this patch is given by $\tilde{P}(N,\tilde{\varphi})$.
After the horizon exit, the averaged AD field obeys the classical equation of motion.
Thus, the condition of $\tilde{\varphi} > \tilde{\varphi_c} \equiv 2\pi\varphi_c/H_I = 2\pi\Delta^{\frac{1}{2}}$ at the end of inflation $N_\mathrm{end}$ can be translated to the condition at the horizon exit that the field value must exceed $\tilde{\varphi}_{\rm{c,eff}}(N)\equiv\exp[c_I'(N_{\rm{end}}-N)/2]\tilde{\varphi}_c$ for the patch that exits the horizon at $N$.

Now, we estimate the volume fraction of HBBs with scales \textit{larger than} $2\pi/k$.
This corresponds to the probability that $\tilde{\varphi}$ exceeds $\tilde{\varphi}_{c,\rm{eff}}(N_{k})$ at the horizon exit.
Thus, it is given by 
\begin{equation}
    B(N_{k})
    =
    \int_{\tilde{\varphi}_{\rm{c,eff}}(N_{k})}^{\infty} \mathrm{d}\tilde{\varphi} \,
    \tilde{P}(N_{k},\tilde{\varphi})
    =
    \exp\left[
        -\frac{\tilde{\varphi}^2_{\rm{c,eff}}(N_{k})}{2\tilde{\sigma}^2(N_{k})}
    \right]
    .
\end{equation}
Then, we can estimate the volume fraction of the HBBs on scales \textit{equal to} $2\pi/k$ by differentiating $B$ with respect to $N_{k}$ as
\begin{equation}
  \beta_{N}(N_{k})
  =
  \frac{\partial B(N_{k})}{\partial N_{k}}
  .
\end{equation}
Considering the relation $M\propto e^{-2N}$, which is derived from Eqs.~\eqref{Hubble mass with respect to k} and \eqref{e-fold number with respect to k}, the PBH formation rate with respect to the logarithm of the PBH mass $\ln M$ can be written as
\begin{equation}
    \beta(M)
    =
    \frac{1}{2}\beta_{N}(N)\theta(M-M_c)
    .
\end{equation}
Here the step function represents that only HBBs with mass larger than $M_c$ collapse into PBHs.
The present energy ratio of PBHs to the total dark matter is estimated by
\begin{equation}
    f_{\rm{PBH}}
    \equiv
    \frac{\Omega_{\rm{PBH}}}{\Omega_c}
    =
    \int \mathrm{d}(\ln M) \,
    \beta(M)\frac{T(M)}{T_{\rm{eq}}}\frac{\Omega_m}{\Omega_c}
    ,
\end{equation}
where $T_{\rm{eq}}$ is the temperature at matter-radiation equality, and $\Omega_m$ and $\Omega_c$ are the density parameters of non-relativistic matter and dark matter, respectively.

\begin{figure}[htbp]
    \centering
    \hspace{-2cm}\includegraphics[width=0.6\linewidth]{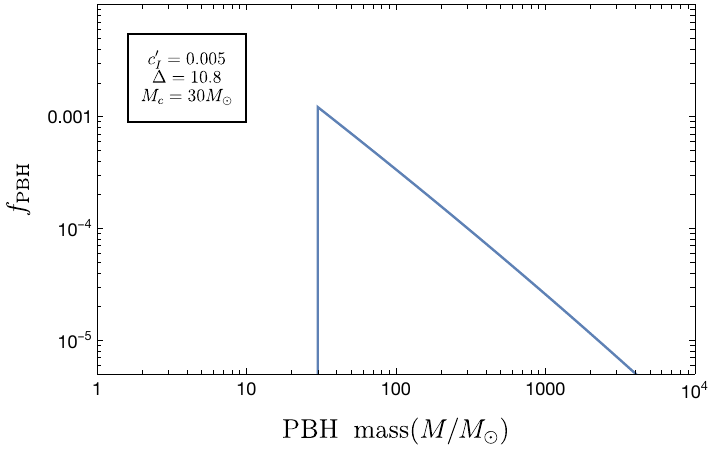}\\
    \includegraphics[width=0.72\linewidth]{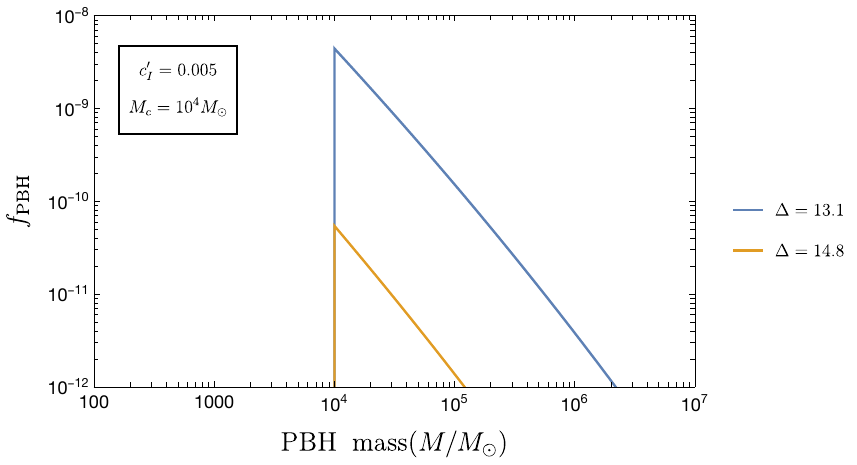}
    \caption{
        PBH mass spectra for $M_c=30M_{\odot}$ to explain the LIGO-Virgo events (upper panel) and for $M_c=10^4M_{\odot}$ to explain the SMBHs above $10^6M_{\odot}$ (lower panel).
        We take $(c_I', \Delta) = (0.005, 10.8)$ (upper blue),  $(c_I', \Delta)=(0.005, 13.1)$ (lower blue) and $(0.005, 14.8)$ (lower orange), which lead to the total abundances of PBH $f_{\rm{PBH}}=1.1\times10^{-3}$, $f_{\rm{PBH}}=3.1\times10^{-9}$ and $3.5\times10^{-11}$, respectively.
    }
    \label{fig:distribution of PBH}
\end{figure}

The PBH mass spectrum is shown in Fig.~\ref{fig:distribution of PBH}.
We can see that there is a steep peak at the threshold mass $M=M_c$ in the PBH mass spectrum.
In the left panel of Fig.~\ref{fig:distribution of PBH}, we choose the parameters
to obtain $M_c=30M_{\odot}$ and $f_{\rm{PBH}} \sim 10^{-3}$~\cite{Sasaki:2016jop}, which is a typical value to explain the LIGO-Virgo events by PBHs assuming that PBHs are not clustered.%
\footnote{
    Although this model predicts a clustering of PBHs~\cite{Kawasaki:2021zir}, it is difficult to evaluate the merger rate for clustered BHs due to the three-body or many-body problem.
    Therefore, we use the PBH abundance that accounts for the LIGO-Virgo events without PBH clustering as a typical value.
}

We also consider the case where the PBHs are seeds of the SMBHs at the center of galaxies.
As we mentioned in Sec.~\ref{sec: intro}, the SMBHs with $M \gtrsim 10^9M_{\odot}$ have been observed at redshift $z\gtrsim 6$. 
The observed mass distribution is consistent with a mass spectrum given by the so-called Schechter function, which gives the SMBH density about $96M_{\odot}~{\rm{Mpc}}^{-3}$ above $M_{\rm{SMBH}}\sim 10^6M_{\odot}$~\cite{Willott:2010yu}.
This mass density corresponds to the DM fraction $f_{\rm{SMBH}}\simeq 2.9\times10^{-9}$.
Now we determine $M_c$ that explains SMBH with masses above $10^6M_{\odot}$.
The numerical simulation shows that PBHs with $M\sim(10^4\textrm{-} 10^5)M_{\odot}$ subsequently grow up to $M\sim 10^9M_{\odot}$~\cite{10.1093/mnras/stw1679}.
However, due to the theoretical uncertainty of the accretion mechanism, the correspondence between $M_{\rm{PBH}}$ and $M_{\rm{SMBH}}$ is unclear.
Here, we assume PBHs with $M_{\rm{PBH}}\gtrsim 10^4M_{\odot}$ become seeds of SMBHs with $M_{\rm{SMBH}}\gtrsim 10^6M_{\odot}$.
Then, we examine the case with $M_c=10^4M_{\odot}$.
In the following, we consider two choices of the PBH abundance.
One is $f_{\rm{PBH}}\simeq 3\times10^{-9}$ assuming that $f_{\rm{PBH}}$ at the PBH formation epoch is equal to $f_{\rm{SMBH}}$ at $z\sim 6$, which may be an overestimation considering accretion.
The other is $f_{\rm{PBH}}\simeq 3\times10^{-11}$ assuming that $f_{\rm{SMBH}}$ at $z\sim 6$ is greater than $f_{\rm{PBH}}$ in the PBH formation epoch by a factor of $10^{2}$ due to accretion.
Note that these PBH abundances are consistent with the CMB constraint from PBH accretion~\cite{Serpico:2020ehh}.

Before closing this section, we mention HBBs with mass smaller than $M_c$.
Such small HBBs do not form PBHs and remain as HBBs, which cause cosmological difficulties as will be seen in the next section.

The total volume fraction of HBBs that do not collapse to PBHs is given by
\begin{equation}
    \beta_{\rm{HBB}}
    \equiv
    \int_{N_c}^{N_{\rm{end}}} \mathrm{d}N \, \beta_{N}(N)
    ,
\end{equation}
where $N_c$ is the e-folding number corresponding to the lower bound of the horizon mass for the PBH formation $M_c$.

\section{BBN constraint on HBB scenario}
\label{sec3}

In order to explain the LIGO-Virgo GW events, we need $\eta^{\rm{in}}_b\sim1$ which gives the PBH threshold mass of $M_c = 30M_{\odot}$ in the case without the stable Q-ball formation.
In the case of the stable Q-ball formation, $\eta^{\rm{in}}_b\sim10^{-2}$ is required to obtain $M_c = 30M_{\odot}$ when we set a typical gravitino mass $m_{3/2}=1~{\rm{GeV}}$.
On the other hand, in order to obtain $M_c = 10^4M_{\odot}$, we require $\eta^{\rm{in}}_b\sim10^{-2}$ in the case without the stable Q-ball formation and $\eta^{\rm{in}}_b\sim10^{-3}$ in the case with the stable Q-ball formation when we set $m_{3/2}=1~{\rm{GeV}}$.
As mentioned above, HBBs smaller than the threshold scale do not collapse into PBHs and the baryon charge in the surviving HBBs can affect BBN.
In this section, we show that the scenario where HBBs collapse into PBHs cannot account for the LIGO-Virgo events or the origin of the SMBHs without spoiling the success of BBN in the case without the stable Q-ball formation or overproducing the dark matter in the case with the stable Q-ball formation.

Each HBB that does not collapse into a PBH has a large baryon asymmetry.
If such HBBs have the same sign of baryon asymmetry, the baryon asymmetry averaged over the observable universe is estimated by
\begin{equation}
    \label{eq:average_baryon}
    \langle \eta_b^{\rm{HBB}}\rangle
    \simeq
    \eta_b^{\rm{in}}
    \int_{N_c}^{N_{\rm{end}}}\beta_{N}(N)\,\mathrm{d}N
    = \eta_b^{\rm{in}}\beta_\text{HBB} .
\end{equation}
In the case without the stable Q-balls, this baryon asymmetry directly affect the BBN and significantly changes the abundances of light elements if $\langle \eta_b^{\rm{HBB}}\rangle$ is larger than the observed baryon asymmetry of the universe $\eta_b^{\rm{obs}}\sim 10^{-10}$.

On the other hand, the baryons in the Q-balls do not participate in the BBN processes in the case with the stable Q-balls.
Instead, the stable Q-balls give a significant contribution to the dark matter density.
The averaged ratio of the Q-ball density to the entropy is written as
\begin{align}
    \langle Y_Q^{\rm{HBB}}\rangle  
    \simeq
    Y_Q^{\rm{in}}
    \int_{N_c}^{N_{\rm{end}}}\beta_{N}(N)\,\mathrm{d}N
    = Y_Q^{\rm{in}}\beta_\text{HBB} = \omega_Q \langle \eta_b^{\rm{HBB}}\rangle ,
\end{align}
which should be smaller than the present dark matter density $Y_\text{DM}^\text{obs}$ ($\sim 2\times 10^{-9}$~GeV).

It is found that $\langle \eta_b^{\rm{HBB}}\rangle$ or $\langle Y_Q^{\rm{HBB}}\rangle$ is much larger than  $\eta_b^{\rm{obs}}$ or $Y_\text{DM}^\text{obs}$ in the parameter region favored to explain the observed BHs with $M\sim 30M_{\odot}$ and SMBH with $M\gtrsim 10^6M_{\odot}$.
Here, we consider the parameter sets in Fig.~\ref{fig:distribution of PBH} and evaluate $\langle \eta_b^{\rm{HBB}}\rangle$ as examples.
For the parameters in the left panel, we obtain the volume fraction of HBBs as $\beta_{\rm{HBB}}\simeq2.7\times10^{-4}$.
Then, the resulting baryon asymmetry becomes $\langle \eta^{\rm{HBB}}_b\rangle \sim 10^{-4} \gg \eta_b^{\rm{obs}}$ in the case without the stable Q-ball formation. In the case with the stable Q-ball formation, $\langle Y_Q^{\rm{HBB}}\rangle \sim 10^{-6}$~GeV $\gg Y_\text{DM}^\text{obs}$.
For the parameters in the right panel with $f_\mathrm{PBH} \simeq 3.5\times 10^{-11}$, $\beta_{\rm{HBB}}\simeq1.3\times10^{-5}$ and the resulting baryon asymmetry becomes $\langle \eta^{\rm{HBB}}_b\rangle \sim 10^{-7} \gg \eta_b^{\rm{obs}}$ in the case without the stable Q-ball formation and $\langle Y_Q^{\rm{HBB}}\rangle \sim 10^{-8}~\mathrm{GeV}\gg Y_\text{DM}^\text{obs}$ in the case with the stable Q-ball formation.
In all of the above cases, $\langle \eta^{\rm{HBB}}_b\rangle$ or $\langle Y_Q^{\rm{HBB}}\rangle$ is much larger than the observed value. 

In fact, the AD mechanism can generate both positive and negative baryon charges depending on the initial phase of the AD field at the onset of oscillations.
Thus, in the case without the stable Q-balls, we can avoid the above problem if the nucleons and anti-nucleons in HBBs with positive and negative baryon charge diffuse and annihilate before BBN.
In order to evaluate this effect, we compare the typical correlation length of the sign of the baryon charge in HBBs with the typical diffusion length of nucleons before BBN.

First, we evaluate the correlation length of the sign of the baryon charge in HBBs.
To this end, we consider two spatial points $x$ and $y$ separated by a comoving distance $l$.
These two points reside in the same Hubble patch until the scale $l$ exits the horizon at $N = N_l$.
Thus, the coarse-grained AD field around these two points is considered to take the same value until $N = N_l$.
After that, the coarse-grained AD field around each of $x$ and $y$ evolves independently.
This evolution is described by Eq.~\eqref{Fokker Planck equation} with the initial condition $P(N_l,\phi) = \delta^{(2)}(\phi-\phi_l)$.
Since we are interested in $c'_I\ll 1$, we ignore the potential term in Eq.~\eqref{Fokker Planck equation} and only consider the quantum fluctuation effect in the following.
We then obtain the probability distribution of $\phi$ for $N > N_l$ as
\begin{align}
    P(N-N_l,\phi;\phi_l) 
    = 
    \frac{A}{N-N_l}
    \exp\left[
        -\frac{4\pi^2|\phi-\phi_l|^2}{H_I^2(N-N_l)}
    \right]
    ,
\end{align} 
where $A$ is a normalization constant.

In the AD mechanism, the sign of the baryon charge in each HBB is determined by the phase of $\phi$ at the end of inflation.

\begin{figure}[t]
    \centering
    \includegraphics[width=0.4\linewidth]{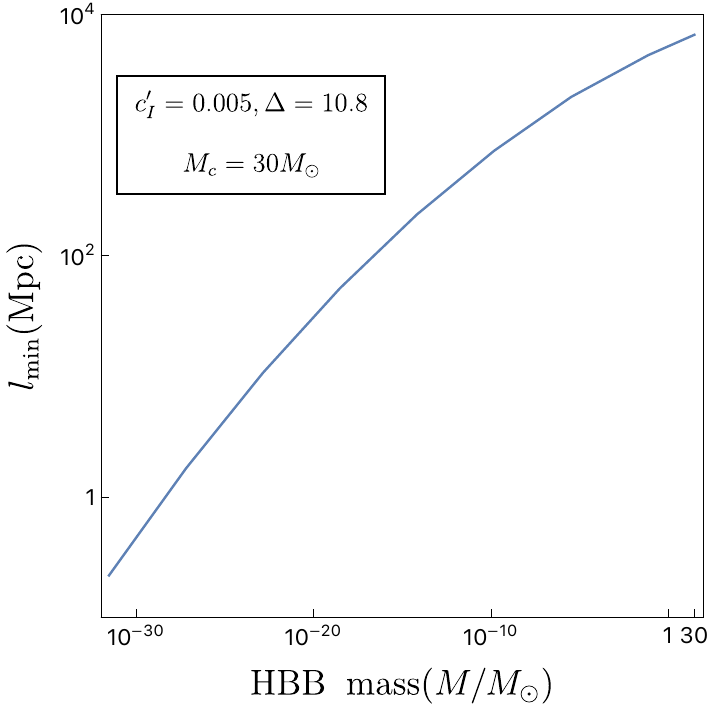}
    \includegraphics[width=0.4\linewidth]{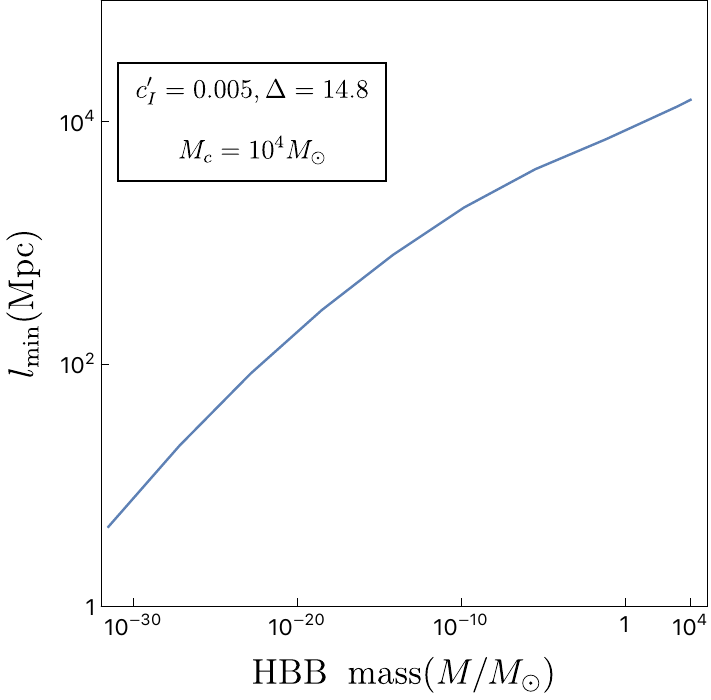}
    \caption{Numerical results of $l_{\rm{min}}$ as a function of the mass of two HBBs for $(c_I',\Delta,M_c)=(0.005,10.8,30M_{\odot})$ (left panel) and $(c_I',\Delta,M_c)=(0.005,14.8,10^4M_{\odot})$ (right panel). 
    One can see that $l_{\rm{min}}\gg l_{\rm{diffuse}}\sim 10^{-11}~{\rm{Mpc}}$ for any value of the HBB mass.
    }
    \label{fig: lmin}
\end{figure}

Thus, we consider the diffusion in the phase direction of the AD field.
In particular, we evaluate the typical difference in the phase of the coarse-grained AD field between two patches that exit the horizon at $N = N_r$ respectively and are separated by a comoving distance of $l$.
We denote the e-folding number corresponding to the distance $l$ by $N_l$.
Since the two patches lie in the same Hubble patch until $N = N_l$, the time evolution of the coarse-grained AD field in the two patches can be regarded as the same.
On the other hand, for $N_l < N < N_r$, the AD field in each patch evolves independently.
Now we are interested in the probability distribution of the difference in the phase of the AD field under the condition that both of the two patches become HBBs.
Since the probability distribution of $\phi$ rapidly decreases for large $\varphi$, we approximate $\varphi$ in each patch at $N = N_r$ by the most likely value for the HBB formation, $\varphi_{c,\mathrm{eff}}(N_r)$.
Then, the probability distribution for the phase difference $\Delta \theta$ at the horizon exit and the coarse-grained AD field value $\phi_l$ at $N = N_l$ is given by 
\begin{align}
    F(\Delta\theta,\phi_l;N_r,N_l)
    \equiv
    P(N_l,\phi_l;0)
    P(N_r-N_l,\varphi_{\rm{c,eff}}(N_r) e^{i\Delta\theta};\phi_l)
    P(N_r-N_l,\varphi_{\rm{c,eff}}(N_r);\phi_l).
\end{align}
Thus, the expectation value of the phase difference at $N = N_r$ between two patches separated by $l$ is estimated as
\begin{align}
    \langle\Delta\theta\rangle(N_r,N_l)
    =
    \frac{
    \int_{0}^{\pi}
    \mathrm{d}(\Delta\theta)
    \int {\mathrm{d}}^2\phi_l\,
    \Delta\theta \cdot F(\Delta\theta,\phi_l;N_r,N_l)
    }
    {
    \int_{0}^{\pi}
    {\mathrm{d}}(\Delta\theta)
    \int {\mathrm{d}}^2\phi_l\, F(\Delta\theta,\phi_l;N_r,N_l)
    }
    ,
\end{align}
where we denote the two-dimensional integral in the complex field space with respect to $\phi_l$ as $\int{\mathrm{d}}^2\phi_l$.
Since the averaged phase of the AD field over the comoving patch is constant after the horizon exit and the asymmetry is generated by the A-term proportional to $\phi^n$, the necessary condition that the signs of the baryon charge in the two HBBs have no correlation can be roughly written as
\begin{equation}
   \langle\Delta\theta\rangle(N_r,N_l)
    \gtrsim
    \frac{\pi}{n}
    .
\end{equation}
This condition gives an inequality for $N_l$ with fixed $N_r$.
Taking into account the relation~\eqref{e-fold number with respect to k} between $k$ and $N$, this gives the minimum distance $l_{\rm{min}}$ between two HBBs with no correlation in the sign of the baryon charge.
We show the numerical results of $l_{\rm{min}}$ in Fig.~\ref{fig: lmin}.

On the other hand, the diffusion length of nucleons before BBN is estimated to be $l_{\rm{diffuse}}\sim10^{-11}~\mathrm{Mpc}$~\cite{Applegate:1987hm}.
The necessary condition for the baryon charges among different HBBs to be canceled by the diffusion process is given by
\begin{equation}
    l_{\rm{min}}(N_r)
    \lesssim
    l_\mathrm{diffuse}
    \quad
    \mathrm{for}\ \ N_c < N_r < N_{\rm{end}}
    \label{constraint by proton diffusion length}
    .
\end{equation}
In Fig.~\ref{fig: lmin}, we can see that $l_{\rm{min}}(N_r)\gg l_{\rm{diffuse}}$ for $N_c<N_r<N_{\rm{end}}$
both in the case of $M_c \simeq 30 M_\odot$ and $10^4 M_\odot$.%
\footnote{
    When we take a larger value of $c_I'$ to explain the same value of $f_\mathrm{PBH}$, the threshold field value $\tilde{\varphi}_{c,\mathrm{eff}}$ becomes smaller and thus $\langle \Delta \theta \rangle$ becomes larger.
    However, the diffusion is still too small to solve the problem of the baryon asymmetry for any value of $c_I'<0.05$, with which we can explain the fixed values of $f_{\rm{PBH}}$ satisfying the requirement of $\Delta\gtrsim 1$.
}
Thus, nearly all baryon charges confined in residual HBBs survive at the BBN epoch, and the light elements are synthesized from the nucleons and anti-nucleons with highly inhomogeneous distribution, which spoils the success of BBN. 
Therefore, the HBB-collapsing scenario cannot account for the LIGO-Virgo observations or SMBHs.
This conclusion does not change if we consider PBHs with other masses in $30M_\odot \leq M_c \leq 10^4M_\odot$.

\section{L-ball scenario}
\label{sec4}

As discussed in the previous section, the PBH formation scenario using a flat direction carrying a baryon number generates a local baryon asymmetry much larger than the observed value $\eta_{\rm{obs}}\sim10^{-10}$, which is disfavored by BBN or the observed dark matter density.
Thus, we consider an alternative scenario using another flat direction carrying a lepton number such as the $LL\bar{e}$ direction.
In this scenario, the generated lepton asymmetry is converted to baryon asymmetry through the anomalous sphaleron processes at the temperature above the electroweak scale.
Therefore, the generated lepton asymmetry should be protected from the sphaleron processes for some reason.
Here, we also assume that the AD field forms Q-balls with a large lepton charge, ``L-balls'', in which the sphaleron processes are not efficient.
However, L-balls gradually evaporate and emit their lepton asymmetry due to the interaction with the thermal plasma.
Moreover, L-balls also decay into standard model particles through the $\phi\phi\rightarrow ll$ process.
The emitted lepton asymmetry is partially converted to baryon numbers by the sphaleron processes.
In the following, we estimate the amount of baryon numbers from these processes and show that there is a parameter region where the generated baryon asymmetry is sufficiently small to have a negligible effect on the baryon asymmetry of the universe. 
We also evaluate the cosmological effect of non-thermal neutrinos emitted from  the L-balls.

\subsection{Evaporation of L-balls}

L-balls gradually emit their charge due to the thermal effect~\cite{Laine:1998rg}.
Here, we estimate the evaporation process following the argument in Ref.~\cite{Kasuya:2014ofa}. 

The evaporation rate is determined by the difference in the chemical potential inside and outside an L-ball, which is given by
\begin{equation}
    \Gamma_{\rm{evap}}
    \equiv
    \left.
        \frac{\mathrm{d}Q}{\mathrm{d}t}
    \right|_{\rm{evap}}
    \sim 
    -4\pi\xi (\mu_Q-\mu_{\rm{plasma}}) T^2R_Q^2
    ,
    \label{eq: evaporation rate}
\end{equation}
where $R_Q$ is the radius of the L-ball, $Q$ is the charge inside the L-ball, and
\begin{equation}
    \xi=\left\{
        \begin{array}{ll}
            1
            \quad&
            (T>m_\phi)
            \\
            \left( \frac{T}{m_\phi} \right)^2
            \quad&
            (T<m_\phi)
        \end{array}
    \right.
    .
    \label{eq:xi}
\end{equation}
Here, $\mu_Q=\omega_Q$ is the chemical potential inside the L-ball, where $\omega_Q$ is the energy per unit charge inside the L-ball.
In the following, we assume that $\mu_Q$ is much larger than the chemical potential of the lepton number in the background plasma, $\mu_{\rm{plasma}}$.
In order for evaporation to proceed efficiently, the evaporated charge should be transferred far from the the surface of the L-ball through diffusion~\cite{Banerjee:2000mb}.
The diffusion rate is evaluated as
\begin{equation}
\Gamma_{\rm{diff}}
    \equiv
    \left.
        \frac{\mathrm{d}Q}{\mathrm{d}t}
    \right|_{\rm{diff}}
    \sim
    -4\pi D R_Q\mu_Q T^2
    ,
    \label{eq: diffusion rate}
\end{equation}
where $D \equiv A/T$ is the diffusion constant and $A=4-6$ is a numerical constant. If the diffusion is inefficient, the evaporation rate is bounded by the diffusion rate.
Thus, we consider that the evaporation rate is given by the smaller of $\Gamma_{\rm{diff}}$ and $\Gamma_{\rm{evap}}$.

\subsection{New-type L-ball scenario}
\label{New type L-ball scenario}

In the gauge-mediated SUSY breaking scenario, the scalar potential including the 2-loop contribution can be written as~\cite{deGouvea:1997afu}
\begin{equation}
    V_{\rm{gauge}}(\phi)
    =
    \left\{
        \begin{array}{ll}
            m_{\phi}^2|\phi|^2
            \quad&
            (|\phi|\ll M_S) 
            \\
            M_F^4 \left( \ln \frac{|\phi|^2}{M_S^2} \right)^2
            \quad&
            (|\phi|\gg M_S)
        \end{array}
    \right.
    ,
    \label{eq:Vgauge}
\end{equation}
where $M_F$ is the SUSY breaking scale and $M_S$ is the mass scale of the messenger sector.
The gravity mediation also contributes to the scalar potential as~\cite{deGouvea:1997afu}
\begin{equation}
    V_{\rm{grav}}(\phi)
    =
    m_{3/2}^2\left[1+K\ln\left(\frac{|\phi|^2}{M_{\rm{Pl}}}\right)\right]|\phi|^2
    ,
\label{eq:Vgrav}
\end{equation}
where $K$ is a dimensionless constant, which we assume to be negative in this section.
L-balls are formed when the potential is dominated by $V_\text{gauge}$.
In this case, the formed L-balls are called ``gauge-mediation-type'' L-balls.
On the other hand, if $V_\text{grav}$ dominates and $K <0$, L-balls called ``new-type'' L-balls are formed~\cite{Kasuya:2000sc}.
This happens for $\varphi \gtrsim \varphi_{\rm{eq}}\equiv M_F^2 / m_{3/2} \gg M_S$.
Here we do not include the thermal potential $V_\text{T}$ because it is almost subdominant for a parameter region that we are interested in.\footnote{
    In the case of the new-type L-ball scenario, the condition that $V_\text{T}$ is subdominant excludes some parameter regions with high reheating temperature.
    However, such a high reheating temperature is already excluded by the constraint from the gravitino overproduction.
}

When the SUSY breaking term or the thermal potential starts to dominate over the negative Hubble induced term in vacuum B, the AD field in vacuum B starts to oscillate around the origin.
The Hubble parameter at the onset of oscillations is obtained by
\begin{align}
    H_{\rm{osc}}
    \simeq
    {\rm{max}}
    \left[
        m_{3/2},~
        \left(\lambda\frac{M_F^{2n-4}}{M_{\rm{Pl}}^{n-3}}\right)^{\frac{1}{n-1}}
    \right]
    ,
\end{align}
where we used the radial component of the AD field in vacuum B given by
\begin{equation}
    \varphi_{\rm{osc}}
    \simeq
    \left(
    \frac{H_{\rm{osc}}M_{\rm{Pl}}^{n-3}}{\lambda}
    \right)^{\frac{1}{n-2}}
    .
    \label{varphi_osc}
\end{equation}
Here, $H_{\rm{osc}}$ is equal to $m_{3/2}$ for typical parameter values. In fact, when we fix the peak mass of the PBH distribution to the observationally motivated values such as $M_c=30M_{\odot}$ or $10^4M_{\odot}$, the inequality $\varphi_{\rm{osc}}>\varphi_{\rm{eq}}$ is satisfied in typical parameter value.
Therefore, we assume that $V_{\rm{grav}}$ is dominant over $V_{\rm{gauge}}$ at the onset of oscillations in the following.
In this case, new-type L-balls are formed by $V_{\rm{grav}}$ since we have assumed $K < 0$.
The lepton charge of a single L-ball is given by
\begin{equation}
    Q_{\rm{N}}
    \simeq
    \beta_N
    \left( \frac{\varphi_{\rm{osc}}}{m_{3/2}} \right)^2
    ,
    \label{New type Lball charge}
\end{equation}
where $\beta_N\sim 0.02$ is a numerical constant~\cite{Hiramatsu:2010dx}.
The mass, radius, and energy per charge of a new-type L-ball are written as
\begin{align}
    \label{eq:mass_new}
    M_Q & \simeq m_{3/2}Q_N,
    \\
    \label{eq:R_new}
    R_Q & \simeq |K|^{-\frac{1}{2}}m_{3/2}^{-1},
    \\
    \label{eq:omega_new}
    \omega_Q & \simeq m_{3/2}.
\end{align}
Substituting these quantities to Eqs.~\eqref{eq: evaporation rate} and \eqref{eq: diffusion rate}, we obtain
\begin{align}
    \label{eq:evap_new}
    \Gamma_{\rm{diff}} &  \simeq -4\pi A|K|^{-\frac{1}{2}}T,
    \\[0.4em]
    \label{eq:diff_new}
    \Gamma_{\rm{evap}} &  \simeq -4\pi\xi |K|^{-1}\frac{T^2}{m_{3/2}}.
\end{align}
$\Gamma_{\rm{diff}}$ and $\Gamma_{\rm{evap}}$ become equal when the temperature becomes $T\sim T_{\ast}$ defined by
\begin{equation}
    T_{\ast}
    \simeq
    \left( A|K|^{\frac{1}{2}} \right)^{\frac{1}{3}}
    \left( m_\phi^2m_{3/2} \right)^{\frac{1}{3}}
    .
\end{equation}
Here, we have used $\xi= T^2/m_\phi^2$, which is justified in the gauge-mediated SUSY breaking model ($m_\phi \gg m_{3/2}$).
Before further calculation, we evaluate the temperature dependence of the physical time as
\begin{equation}
    \frac{\mathrm{d}t}{\mathrm{d}T}
    =
    \begin{cases}
        -\frac{8}{3}M_{\rm{Pl}}T_{\rm{RH}}^2T^{-5}
        \quad&
        (T\gtrsim T_{\rm{RH}})
        \\[0.6em]
        -\frac{3}{\pi}
        \sqrt{\frac{10}{g_{\ast}}}
        M_{\rm{Pl}}T^{-3}
        \quad&
        (T\lesssim T_{\rm{RH}})
    \end{cases}
    ,
    \label{temperature-time}
\end{equation}
where $g_{\ast}$ is the total relativistic degrees of freedom.

Now we calculate the evaporated charge in the two different situations: (i) $T_{\rm{RH}}>T_{\ast}$ and (ii) $T_{\rm{RH}}<T_{\ast}$.
We assume $T_{\ast}>T_{\rm{ew}}$ in both cases, where $T_{\rm{ew}}$ ($\sim 100$~GeV) is the electroweak scale.
First, we consider case (i): $T_{\rm{RH}}>T_{\ast}$.
Taking Eqs.~\eqref{eq:evap_new}, \eqref{eq:diff_new}, and \eqref{temperature-time} into account, the derivative of the L-ball charge with respect to $T$ is obtained as
\begin{equation}
    \frac{\mathrm{d}Q}{\mathrm{d}T}
    \simeq
    \left\{
        \begin{array}{ll}
            \frac{32}{3}\pi A|K|^{-\frac{1}{2}}M_{\rm{Pl}}T_{\rm{RH}}^2T^{-4}
            \quad &
            (T\gtrsim T_{\rm{RH}})
            \\[0.6em]
            12A\sqrt{\frac{10}{g_{\ast}}}
            |K|^{-\frac{1}{2}}M_{\rm{Pl}}T^{-2}
            \quad &
            (T_{\ast}\lesssim T\lesssim T_{\rm{RH}})
            \\[0.6em]
            12|K|^{-1}
            \sqrt{\frac{10}{g_{\ast}}}
            \frac{M_{\rm{Pl}}T}{m_\phi^2m_{3/2}}
            \quad&
            (T\lesssim T_{\ast})
        \end{array}
    \right.
    .
\end{equation}
We estimate the total evaporated lepton charge by integrating these formulae from $T_{\rm{ew}}$ to $T_{\rm{osc}}$.
Moreover, using the relation $n_B|_{\rm{eq}}=\frac{8}{23} (n_B-n_L)|_{\rm{initial}}$ in the chemical equilibrium in the two Higgs doublet model, we obtain the total baryon charge from an L-ball as
\begin{align}
    \Delta Q_b^{\rm{evap}}
    \simeq 
    &\frac{256}{207}\pi A|K|^{-\frac{1}{2}}\frac{M_{\rm{Pl}}}{T_{\rm{RH}}}
    \left(
        1-\frac{27}{8\pi}\sqrt{\frac{10}{g_{\ast}}}
    \right)
    +\frac{144}{23}A^{\frac{2}{3}}|K|^{-\frac{2}{3}}\sqrt{\frac{10}{g_{\ast}}}
    \frac{M_{\rm{Pl}}}{(m_\phi^2m_{3/2})^{\frac{1}{3}}}
    \nonumber\\
    &-\frac{48}{23}|K|^{-1}\sqrt{\frac{10}{g_\ast}}\frac{M_{\rm{Pl}}T_{\rm{ew}}^2}{m_\phi^2m_{3/2}}
    ,
\end{align}
where we assume $T_{\rm{osc}} \gg T_{\rm{RH}}$.

Next, let us consider case (ii): $T_{\rm{RH}}<T_{\ast}$.
$\text{d}Q/\text{d}T$ is given by
\begin{equation}
    \frac{\mathrm{d}Q}{\mathrm{d}T}
    \sim
    \left\{
        \begin{array}{ll}
            \frac{32}{3}\pi A|K|^{-\frac{1}{2}}M_{\rm{Pl}}T_{\rm{RH}}^2T^{-4}
            \quad&
            (T\gtrsim T_{\ast})
            \\[0.6em]
            \frac{32}{3}\pi|K|^{-1}M_{\rm{Pl}}\frac{T_{\rm{RH}}^2}{m_\phi^2m_{3/2}}T^{-1}
            \quad&
            (T_{\rm{RH}}\lesssim T\lesssim T_{\ast})
            \\[0.6em]
            12|K|^{-1}\left(\frac{10}{g_{\ast}}\right)^{\frac{1}{2}}\frac{M_{\rm{Pl}}T}{m_\phi^2m_{3/2}}
            \quad&
            (T\lesssim T_{\rm{RH}})
        \end{array}
    \right.
    .
\end{equation}
By integrating these formulae, we obtain
\begin{equation}
    \Delta Q_b^{\rm{evap}}
    \sim 
    \frac{256}{207}\pi|K|^{-1}
    \left[
        1 + 3\ln\frac{T_{\ast}}{T_{\rm{RH}}}
        +\frac{27}{16\pi}\sqrt{\frac{10}{g_{\ast}}}
        \left\{
            1-\left(\frac{T_{\rm{ew}}}{T_{\rm{RH}}}\right)^2
        \right\}
    \right]
    \frac{M_{\rm{Pl}}T_{\rm{RH}}^2}{m_\phi^2m_{3/2}}
    ,
\end{equation}
where we assume $T_{\rm{osc}} \gg T_{\ast}$.

Now, we estimate the decay temperature of L-balls.
Since $\omega_Q (\simeq m_{3/2})$ is much smaller than the neutralino mass $m_{\tilde{N}}$, the decay process $\phi \rightarrow \nu + \tilde{N}$ is forbidden kinematically.
We also note that charged lepton cannot be emitted due to electric charge conservation.
The main channel of the decay is $\phi+\phi \rightarrow \nu +\nu$ via gaugino exchange.
Since the decay into fermions inside L-balls is suppressed because of the Pauli blocking, the decay rate is bounded by the outgoing flux of the neutrinos from the surface~\cite{Cohen:1986ct}.
This maximum decay rate is realized in the case of the L-ball decay into neutrinos.
Thus, the decay rate of L-balls is given by~\cite{Kawasaki:2012gk,Kasuya:2012mh}
\begin{align}
    \frac{\mathrm{d}Q}{\mathrm{d}t}
    \simeq
    \frac{\omega_Q^3 R_Q^2}{\pi}
    .
    \label{upper limit of decay process}
\end{align}
Taking Eqs.~\eqref{temperature-time} and~\eqref{upper limit of decay process} into account, the lepton charge emitted through the decay process between $T_\text{osc}$ and $T$ ($T_\text{osc} > T_\text{RH} > T)$ is estimated as 
\begin{align}
    \Delta Q_{\rm{decay}}
    \simeq 
    \frac{2|K|^{-1}}{3\pi} m_{3/2} M_{\rm{Pl}} T_{\rm{RH}}^2
    (T_{\rm{RH}}^{-4}-T_{\rm{osc}}^{-4})
    +
    \frac{3|K|^{-1}}{2\pi^2} \sqrt{\frac{10}{g_{\ast}}}
    m_{3/2}M_{\rm{Pl}}
    (T^{-2}-T_{\rm{RH}}^{-2})
    .
    \label{eq: Qdecay in new-type}
\end{align}
Setting $\Delta Q_{\rm{decay}} \simeq Q_N$ (the initial charge), we obtain the decay temperature as
\begin{equation}
    T_{\rm{decay}}
    \simeq
    \left(
        \frac{3}{2\pi^2}\sqrt{\frac{10}{g_{\ast}}}|K|^{-1}m_{3/2}M_{\rm{Pl}}Q_N^{-1}
    \right)^{\frac{1}{2}}
    ,
    \label{Tdecay}
\end{equation}
where we assumed $T_\text{decay} \ll T_\text{RH} < T_\text{osc}$.

Now we search for a parameter space where the L-ball scenario successfully accounts for the abundances of PBHs with masses $30M_\odot$ and $10^4M_\odot$.
Hereafter, we set $c_M=|a_M|=1$ for simplicity, and set $m_\phi=10^4~{\rm{GeV}}$.
First, we fix the threshold mass $M_c$.
Then, from Eq.~\eqref{threshold mass}, we determine $Y_L^{\rm{in}}$ as a function of $M_c$,
\begin{equation}
    Y_L^{\rm{in}}
    \simeq 
    120~{\mathrm{MeV}}
    \times
    \left( \frac{g_{\ast}}{10.75} \right)^{-\frac{1}{8}}
    \left( \frac{M_c}{M_{\odot}} \right)^{-\frac{1}{2}}
    .
    \label{Y_L_fix}
\end{equation}
Combining this with Eqs.~\eqref{baryon number made from AD mechanism}, \eqref{definition of Y_Q}, and \eqref{varphi_osc}, we can represent $\lambda$ as a function of $m_{3/2}$ and $T_{\rm{RH}}$ for a fixed $M_c$,
\begin{equation}
    \lambda
    \simeq
    2.9\times10^{-13}
    \left(
        \frac{g_{\ast}}{10.75}
    \right)^{1/4}
    \left(
        \frac{M_c}{M_{\odot}}
    \right)
    \left(
        \frac{m_{3/2}}{1~{\rm{GeV}}}
    \right)
    \left( 
        \frac{T_{\rm{RH}}}{10^3~{\rm{GeV}}}
    \right)^2
    .
\end{equation}
Here, we take $n=6$ assuming the situation where the AD field corresponds to the $LL\bar{e}$ direction%
\footnote{
If we choose other leptonic flat directions such as $QL\bar{d}$, the discussion above is almost the same.
In this case, we should take $n=4$ assuming the situation where the direction is lifted by $W=QQQL$ or $W=QuQd$.
}
and is lifted by a superpotential $W= LL\bar{e}LL\bar{e}$.
Using this formula, we can express the quantities related to L-balls as functions of $m_{3/2}$ and $T_{\rm{RH}}$.
From Eq.~\eqref{New type Lball charge}, the initial lepton charge of an L-ball can be written as 
\begin{equation}
    Q_N
    \simeq
    1.4\times10^{32}
    \left(
        \frac{g_{\ast}}{10.75}
    \right)^{-1/8}
    \left(
        \frac{M_c}{M_{\odot}}
    \right)^{-1/2}
    \left(
        \frac{m_{3/2}}{1~{\rm{GeV}}}
    \right)^{-2}
    \left(
        \frac{T_{\rm{RH}}}{10^3~{\rm{GeV}}}
    \right)^{-1}.
    \label{Q_N fixed}
\end{equation}
In addition, using Eqs.~\eqref{Tdecay} and \eqref{Q_N fixed} , $T_{\rm{decay}}$ can be written as
\begin{equation}
    T_{\rm{decay}}
    \simeq 
    0.16~{\rm{keV}}
    \times
    \left(
        \frac{g_{\ast}(T_{\rm{decay}})}{10.75}
    \right)^{1/16}
    \left(
    \frac{|K|}{0.1}
    \right)^{-1/2}
    \left(
        \frac{M_c}{M_{\odot}}
        \right)^{1/4}
    \left(
    \frac{m_{3/2}}{1~{\rm{GeV}}}
    \right)^{3/2}
    \left(
        \frac{T_{\rm{RH}}}{10^3~{\rm{GeV}}}
    \right)^{1/2}.
    \label{Tdecay as function of parameters}
\end{equation}

Since the L-balls are produced highly inhomogeneously, the baryon charge $\Delta \eta_b$ converted from the lepton charge emitted by L-balls is also highly inhomogeneous and cannot explain the observed baryon asymmetry of the universe.
Thus, a large value of $\Delta \eta_b$ may spoil the success of the standard BBN.
In this paper, we require that $\Delta \eta_b$ should be smaller than 1\% of the observed baryon asymmetry not to affect BBN.

In other words, the extra baryon asymmetry averaged over the observable universe $\langle \Delta \eta_b \rangle$ should satisfy
\begin{equation}
    \langle \Delta \eta_b \rangle
    =
    \eta_L^{\rm{in}}
    \frac{\Delta Q_b^{\rm{evap}}
    +
    \Delta Q_b^{\rm{decay}}(T_{\rm{ew}})}{Q_N}
    \beta_\mathrm{HLB}
    \lesssim 
    10^{-2}\,
    \eta_b^{\rm{obs}}
    .
    \label{upper bound of baryon asymmetry}
\end{equation}

Next, we consider the energy density of neutrinos emitted from the L-ball decay.
We require that the contribution of such neutrinos to the effective number of neutrino species $N_\mathrm{eff}$ should not exceed the observational upper bound.
The effective number of neutrino species is defined by
\begin{equation}
    N_{\rm{eff}}
    =
    \frac{8}{7}
    \left(\frac{11}{4}\right)^{\frac{4}{3}}
    \left[\frac{\rho_{\rm{rad}}}{\rho_{\rm{\gamma}}}-1\right]
    ,
\end{equation}
where $\rho_\mathrm{rad}$ is the total energy density of radiation, and $\rho_\gamma$ is the energy density of photons.
Assuming that the L-ball decay occurs instantaneously at $T\simeq T_{\rm{decay}}$, the change of $N_{\rm{eff}}$ due to the L-ball decay is estimated as
\begin{equation}
    \Delta N_{\rm{eff}}
    =
    \frac{8}{7}
    \left(\frac{11}{4}\right)^{\frac{4}{3}}
    \frac{\Delta\rho_\nu}{\rho_{\rm{\gamma}}}
    =
    \frac{8}{7}
    \left(\frac{11}{4}\right)^{\frac{4}{3}}
    \frac{Y_L^{\rm{in}}
    s_{\rm{tot}}(T_{\rm{decay}})}
    {\rho_{\rm{\gamma}}(T_{\rm{decay}})}
    \beta_\mathrm{HLB}
    ,
    \label{Delta_N_fix}
\end{equation}
where $\Delta \rho_\nu$ is the energy density of neutrinos emitted from the L-ball decay, and $s_\mathrm{tot}(T)$ is the total entropy density at $T$. 
We require $\Delta N_{\rm{eff}} < 0.24$ from the Planck 2018 results $N_\mathrm{eff}=2.92^{+0.36}_{-0.37}$ ~\cite{Planck:2018vyg} with the standard value of $N_\mathrm{eff}=3.044$~\cite{Akita:2020szl} taken into account.
Using Eq.~\eqref{Y_L_fix}, we obtain the lower bound on $T_\mathrm{decay}$ as 

\begin{equation}
    T_{\rm{decay}}
    \gtrsim 
    T_\mathrm{d,min}
    \equiv
    15~{\mathrm{GeV}}
    \times
    \left(\frac{M_c}{M_{\odot}}\right)^{-\frac{1}{2}}
    \beta_\mathrm{HLB}
    .
    \label{lower bound of Tdecay}
\end{equation}

Taking Eq.~\eqref{Tdecay as function of parameters} into account, Eq.~\eqref{lower bound of Tdecay} constraints on $m_{3/2}$ and $T_{\rm{RH}}$.

\begin{figure}[t]
    \centering
    \includegraphics[width=0.48\linewidth]{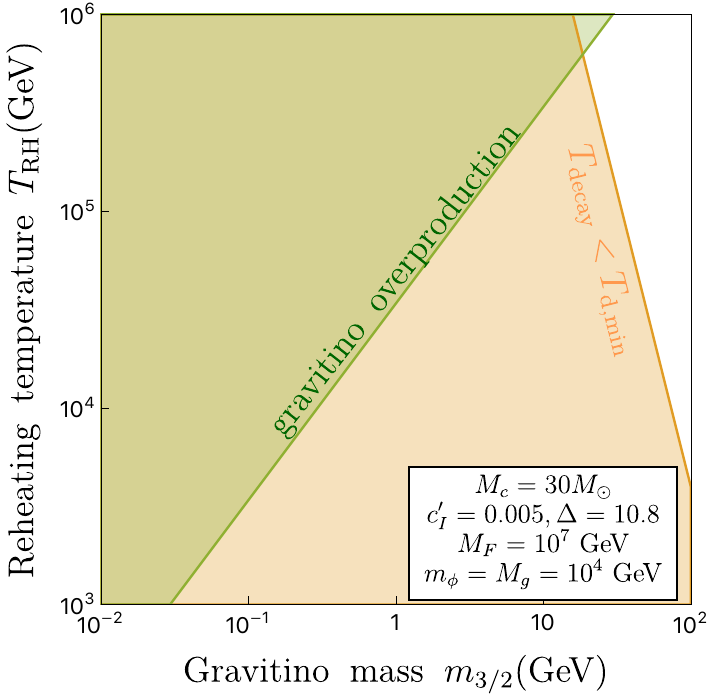}
    \includegraphics[width=0.48\linewidth]{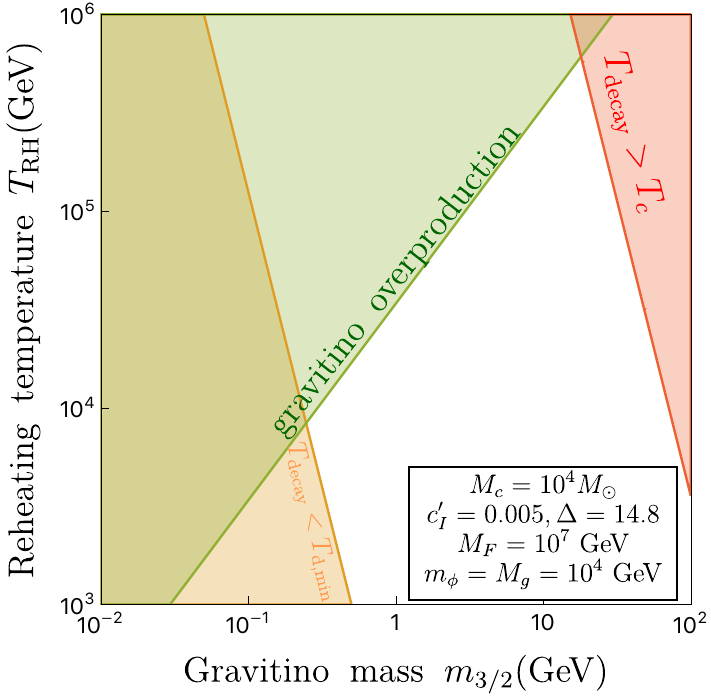}
    \caption{%
    Constraints on $m_{3/2}$ and $T_\mathrm{RH}$ in the new-type L-ball scenario for $M_F=10^7~{\rm{GeV}}$ and $m_\phi=10^4~{\rm{GeV}}$.
    We take $(c_I',\Delta,f_{\rm{PBH}})=(0.005,10.8,1.1\times10^{-3})$ for $M_c=30M_{\odot}$ (left panel) and $(c_I',\Delta,f_{\rm{PBH}})=(0.005,14.8,3.5\times10^{-11})$ for $M_c=10^4M_{\odot}$ (right panel).
    In the green regions, the gravitino density parameter $\Omega_{3/2}$ exceeds the dark matter density $\Omega_{\rm{DM}}\simeq0.12 h^{-2}$ for the gluino mass $M_g=10^4~{\rm{GeV}}$.
    The orange regions are excluded from  the observation of $N_\mathrm{eff}$ [Eq.~\eqref{lower bound of Tdecay}] and the red region in the right panel is excluded because L-balls decay before the PBH formation.
    Any parameters in these panels are consistent with the constraint from the baryon asymmetry in Eq.~\eqref{upper bound of baryon asymmetry}.
    }
    \label{NewType L-ball PBH}
\end{figure}

Finally, we discuss the gravitino problem~\cite{Moroi:1993mb,Kawasaki:2022hvx}.
Gravitinos are produced at the reheating epoch after inflation.
They are stable in gauge-mediated SUSY breaking and give a significant contribution to the matter density.
When the reheating temperature $T_{\rm{RH}}$ is too high, gravitinos are overproduced, which causes a cosmological difficulty called gravitino problem.
The gravitino density parameter $\Omega_{3/2}$ is given by~\cite{Bolz:2000fu,Pradler:2006qh}%
\footnote{
    Although thermal gravitino production is also discussed in Refs.~\cite{Rychkov:2007uq,Eberl:2020fml}, we take the conservative constraint.
    
}
\begin{equation}
    \Omega_{3/2}h^2\simeq 0.71
    \left(
    \frac{m_{3/2}}{0.5~{\rm{GeV}}}
    \right)^{-1}
    \left(
    \frac{M_g}{10^4~{\rm{GeV}}}
    \right)^2
    \left(
    \frac{T_{\rm{RH}}}{10^5~{\rm{GeV}}}
    \right),
\end{equation}
where $M_g$ is the gluino mass and $h$ is the current Hubble parameter in units of 100~km/s/Mpc.
This gravitino abundance should be less than the dark matter density $\Omega_{\rm{DM}}h^2 \simeq 0.12$, which leads to the constraint on $T_{\rm{RH}}$. 
We show the parameter region which is consistent with the above constraints in Fig.~\ref{NewType L-ball PBH}.
One can see that only the region with gravitino mass $m_{3/2}\gtrsim10~{\rm{GeV}}$  and reheating temperature $T_{\rm{RH}}\gtrsim 10^4~{\rm{GeV}}$ is allowed in the case with $M_c=30M_{\odot}$.
Since the L-balls with large lepton charge have long lifetimes if the gravitino mass is small, their density is large at the decay epoch and increases $N_\text{eff}$, which leads to the stringent constraint on $m_{3/2}$.
The required gravitino mass is larger than that favored in the gauge-mediated SUSY breaking scenario ($m_{3/2} \lesssim 1$~GeV).
On the other hand, for $M_c=10^4M_{\odot}$, the region with $m_{3/2}\gtrsim 0.3~{\rm{GeV}}$ and $T_\text{RH} \lesssim 10^6$~GeV is allowed.
In this case, the L-balls have smaller charge because the lepton asymmetry in HLBs decreases as the threshold mass $M_c$ increases [see Eq.~\eqref{Y_L_fix}].
Thus, the L-balls decay earlier.
In addition, the PBH abundance is much smaller than that for the LIGO-Virgo events and hence the L-ball density is smaller.
These two effects weaken the constraint from $N_\text{eff}$ and allow the gravitino mass as small as $\sim 1$~GeV.

\subsection{Delayed-type L-ball Scenario}
\label{Delayed-type L-ball Scenario}

Here, we consider the L-ball scenario in the gauge-mediated SUSY breaking with $K>0$.
In this case, the L-ball solution does not exist as long as $V_{\rm{grav}}$ dominates the total potential.
Thus, when the AD field with an amplitude larger than $\varphi_\text{eq}$ starts to oscillate, L-balls do not form.
When the amplitude decreases by the cosmic expansion and $V_{\rm{grav}} \lesssim V_{\rm{gauge}}$ is satisfied, L-balls are formed.
They are called ``delayed-type'' L-balls~\cite{Kasuya:2001hg,Kawasaki:2002hq,Kasuya:2014ofa}.
Their charge is given by 
\begin{align}
    Q_G & \simeq \beta_G \frac{\varphi_{\rm{eq}}^4}{M_F^4}
    \simeq
    \beta_G 
    \left(
    \frac{M_F}{m_{3/2}}
    \right)^4
    , 
    \label{Q_G}
\end{align}
where $\beta_G\simeq 6\times10^{-4}$ is a numerical constant~\cite{Kasuya:2014ofa}.
Delayed-type L-balls have the same properties as the gauge-mediated-type Q-balls.
Then, the mass, radius, and energy per charge of each delayed-type L-ball are given by
\begin{align}
    M_Q & \simeq \frac{4\sqrt{2}\pi}{3}\zeta M_F Q_G^{\frac{3}{4}}, \\[0.4em]
    R_Q & \simeq \frac{1}{\sqrt{2}\zeta}M_F^{-1}Q_G^{\frac{1}{4}}, \\[0.4em]
    \omega_Q & \simeq \sqrt{2}\pi \zeta M_F Q_G^{-\frac{1}{4}},
\end{align}
where $\zeta =\mathcal{O}(1)$ is a numerical constant. 

The rates of evaporation and diffusion of L-balls are obtained by
\begin{align}
    \Gamma_{\rm{diff}}
    &\simeq
    -4\pi^2AT,
    \\
    \Gamma_{\rm{evap}}
    &\simeq
    -2\sqrt{2}\pi^2\xi\zeta^{-1}\frac{T^2}{m(T)}
    Q_G^{\frac{1}{4}}
    ,
\end{align}
where
\begin{equation}
    m(T)=\left\{
        \begin{array}{ll}
            T \quad& (T>M_F) 
            \\
            M_F \quad& (T<M_F)
        \end{array}
    \right.
    .
\end{equation}
The evaporation rate is equal to the diffusion rate at $T_{\ast}\simeq (\sqrt{2}\zeta A)^{\frac{1}{3}}(m_\phi^2M_F)^{\frac{1}{3}}Q_G^{-\frac{1}{12}}$, which satisfies $T_* < m_\phi < M_F$ for the typical model parameters.

In a similar way to the new-type L-ball case, we obtain the evaporated lepton charge as

\begin{align}
    \Delta Q^{\rm{evap}}_b 
    \simeq &
    \frac{8}{23}
    \left(
    \frac{32}{9}
    \pi^2A\frac{M_{\rm{Pl}}}{T_\mathrm{RH}}\left(1-\frac{27}{8\pi}\sqrt{\frac{10}{g_{\ast}}}\right)
    +
    2^{\frac{5}{6}}9\pi A^{\frac{2}{3}}\sqrt{\frac{10}{g_{\ast}}}\frac{M_{\rm{Pl}}}{(m_\phi^2M_F)^{\frac{1}{3}}}Q_G^{\frac{1}{12}}
    \right.
    \nonumber \\
    &
    \left.
    - 3\sqrt{2}\sqrt{\frac{10}{g_{\ast}}}\pi\frac{M_{\rm{Pl}}T_{\rm{ew}}^2}{m_\phi^2M_F}Q_G^{\frac{1}{4}}
    \right)
    ~~~~~~~~(T_{\ast} \lesssim T_{\rm{RH}}),
    \\[0.5em]
    \Delta Q^{\rm{evap}}_b
    \simeq &
    \left[
        1
        +
        3\ln\frac{T_{\ast}}{T_{\rm{RH}}}
        +
        \frac{27}{16\pi}\sqrt{\frac{10}{g_{\ast}}}\left(1-\left(\frac{T_{\rm{ew}}}{T_{\rm{RH}}}\right)^2\right)
    \right]
    \nonumber \\
    & 
    \times \frac{128\sqrt{2}}{207}\pi^2
    \frac{M_{\rm{Pl}}T_{\rm{RH}}^2}{m_\phi^2M_F}Q_G^{\frac{1}{4}}
    ~~~~~~~~~~~(T_{\ast}\gtrsim T_{\rm{RH}}) .
\end{align}
As for the decay rate, from Eq.~\eqref{upper limit of decay process} we obtain
\begin{equation}
    \left.\frac{\mathrm{d}Q}{\mathrm{d}T}\right|_{\rm{decay}}
    \simeq
    \left\{
    \begin{array}{ll}
    \frac{8\sqrt{2}}{3}\pi^2M_{\rm{Pl}}T_\mathrm{RH}^2M_FQ_{G}^{-\frac{1}{4}}T^{-5}
    & \quad
    (T\gtrsim T_{\rm{RH}}) 
    \\[0.5em]
    3\sqrt{2}\pi M_{\rm{Pl}}M_FQ_{G}^{-\frac{1}{4}}T^{-3}
    & \quad
    (T\lesssim T_{\rm{RH}})
    \end{array}
    \right.
    .
\end{equation}
By integrating these formulae, we obtain the rest charge in the L-ball as 
\begin{equation}
    Q
    \simeq 
    \left[
        \frac{15}{4\sqrt{2}}\sqrt{\frac{10}{g_{\ast}}}\pi M_{\rm{Pl}}M_F(T_\mathrm{RH}^{-2}-T^{-2})
        +\frac{5}{3\sqrt{2}}\pi^2M_{\rm{Pl}}M_FT_\mathrm{RH}^2(T_{\rm{eq}}^{-4}-T_\mathrm{RH}^{-4})
        +Q_G^{\frac{5}{4}}
    \right]^{\frac{4}{5}}
    .
\end{equation}
Since the contribution of the first term is dominant compared to the second term, we obtain the decay temperature as 
\begin{equation}
    T_{\rm{decay}}
    \simeq
    \left(\frac{15}{4\sqrt{2}}\pi M_{\rm{Pl}}M_F Q_G^{-\frac{5}{4}}\right)^{\frac{1}{2}}
    \simeq
    \left(\frac{15}{4\sqrt{2}} \pi\right)^{\frac{1}{2}}\beta_G^{-5/8} M_{\rm{Pl}}^{1/2}
    \frac{m_{3/2}^{5/2}}{M_F^2}
    ,
\end{equation}
where we have used $g_{\ast}\sim10$ and Eq.~\eqref{Q_G}.
For a fixed $M_c$, we write $\lambda$ in terms of $m_{3/2}$, $T_{\rm{RH}}$, and $M_c$ as
\begin{equation}
    \lambda
    \simeq
    2.3\times10^{-10}
    \left(
        \frac{g_{\ast}}{10.75}
    \right)^{1/4}
    \left(
        \frac{M_c}{M_{\odot}}
    \right)
    \left(
        \frac{m_{3/2}}{1~{\rm{GeV}}}
    \right)
    \left( 
        \frac{T_{\rm{RH}}}{10^3~{\rm{GeV}}}
    \right)^2
    ,
\end{equation}
where we fixed $n=6$.
The formulae for the conditions for the baryon density and $N_\mathrm{eff}$ are the same as in the previous section.

There is an additional condition for the delayed-type L-ball scenario.
In this scenario, the AD field starts to oscillate from $\varphi_\mathrm{osc} > \varphi_\mathrm{eq}$, where $V_\mathrm{grav}$ dominates the potential.
After that, delayed-type L-balls are formed when $V_\mathrm{gauge}$ dominates the potential at $\varphi \simeq \varphi_\mathrm{eq}$.
Therefore, we require $\varphi_{\rm{osc}} > \varphi_{\rm{eq}}$ to realize the delay of L-ball formation.

\begin{figure}[t]
\centering
    \begin{tabular}{c}
        \includegraphics[width=0.48\linewidth]{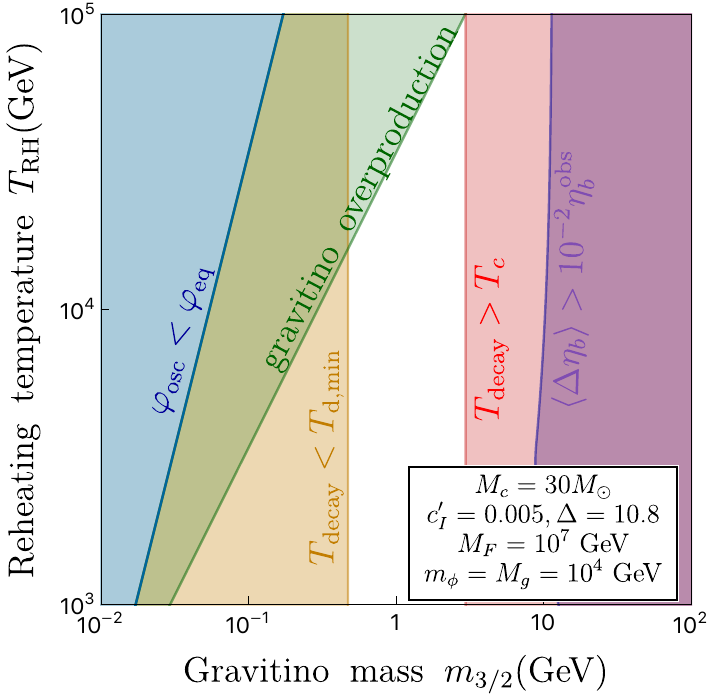}
        \includegraphics[width=0.48\linewidth]{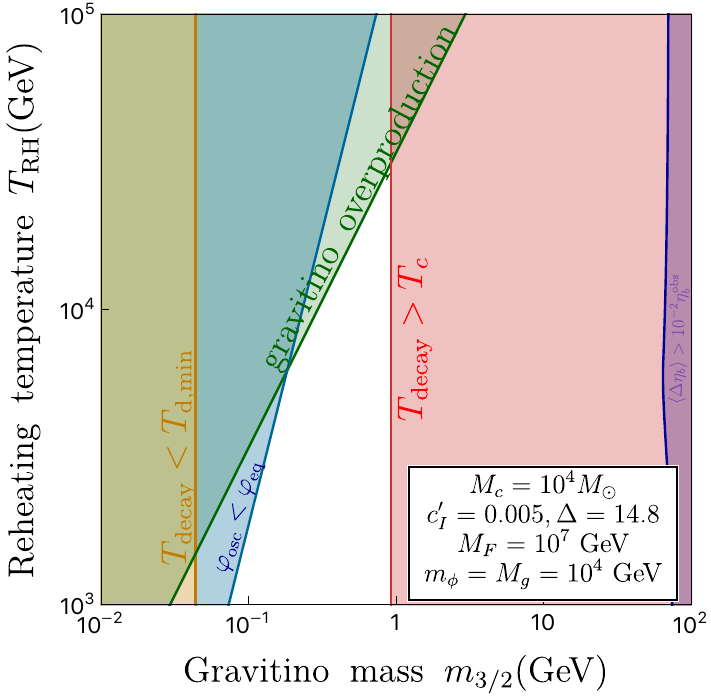}
    \end{tabular}
    \caption{%
    Constraints on $m_{3/2}$ and $T_\mathrm{RH}$ in the delayed-type L-ball scenario for $M_F=10^7~{\rm{GeV}}$ and $m_\phi=10^4~{\rm{GeV}}$.
    We take $(c_I',\Delta,f_{\rm{PBH}})=(0.005,10.8,1.1\times10^{-3})$ for $M_c=30M_{\odot}$ (left panel) and $(c_I',\Delta,f_{\rm{PBH}})=(0.005,14.8,3.5\times10^{-11})$ for $M_c=10^4M_{\odot}$ (right panel).
    The green regions are excluded from the gravitino problem as in Fig.~\ref{NewType L-ball PBH}.
    The orange regions are excluded from the observation of $N_\mathrm{eff}$ [Eq.~\eqref{lower bound of Tdecay}], and the red regions are excluded because the L-balls decay before the PBH formation.
    In the blue regions, our assumption $\varphi_\mathrm{osc} > \varphi_\mathrm{eq}$ is not satisfied.
    In the purple regions, the baryon asymmetry from L-balls can spoil the success of the standard BBN. 
    }
    \label{Delayed L-ball PBH}
\end{figure}

We show the parameter region consistent with the above constraints in Fig.~\ref{Delayed L-ball PBH}.
It is seen that the region around $m_{3/2}\sim 1~{\rm{GeV}}$ is allowed in the case with $M_c=30M_{\odot}$, and the region with $0.1~{\rm{GeV}}\lesssim m_{3/2}\lesssim1~{\rm{GeV}}$ and $T_{\rm{RH}}\lesssim 10^4~{\rm{GeV}}$ is allowed in the case with $M_c=10^4M_{\odot}$.
In the delayed-type L-ball scenario, the formed L-balls have smaller charges than the new-type L-balls because the field amplitude at the formation epoch is smaller for a given lepton asymmetry, which results in earlier decay of the L-balls and hence weaker constraint from $N_\text{eff}$.
On the other hand, since the L-balls decay earlier, the constraint from the requirement that the decay should take place after the PBH formation becomes stringent in the delayed-type L-ball scenario.
Thus, the gravitino mass is determined as $m_{3/2} \sim 1$~GeV by these constraints.

\section{Conclusion}
\label{sec5}

In this paper, we have considered the Affleck-Dine mechanism for the PBH formation, focusing on two PBH mass regions: $M \sim 30 M_\odot$ motivated by the LIGO-Virgo observations of the binary black hole mergers and $M \gtrsim 10^4 M_\odot$ motivated by the observations of supermassive black holes at the center of galaxies.
The mechanism produces high baryon bubbles (HBBs)
that collapse into PBHs if their horizon mass is larger than some critical value.
However, small HBBs that do not collapse into PBHs give a too large contribution to the baryon density at the BBN epoch or the present dark matter density if the stable Q-balls are formed in HBBs. 
We have found that this problem cannot be solved by the existence of HBBs with positive and negative baryon numbers because the correlation length between HBBs with opposite baryon charges is much larger than the nucleon diffusion length, and hence nucleons and anti-nucleons cannot annihilate.

Therefore, we have proposed an alternative scenario where the AD mechanism produces regions with large lepton numbers and Q-balls with lepton charge ($=$ L-balls) are produced therein.
The formed L-balls confine all the produced lepton number inside them and protect it from the sphaleron processes.
However, some fraction of the lepton charge is emitted from the L-balls through the evaporation and decay processes and converted to baryon charge.
Moreover, the L-balls finally decay into neutrinos.
We have estimated the baryon asymmetry and increases of the effective number of neutrino species produced by the evaporation and decay.
It is found that there is a parameter region where the L-ball scenario can account for the LIGO-Virgo GW events and SMBHs. 

\begin{acknowledgments}
This work was supported by JSPS KAKENHI Grant Nos. 20H05851(M.K.), 21K03567(M.K.), JP20J20248 (K.M.) and World Premier International Research Center Initiative (WPI Initiative), MEXT, Japan (M.K., K.M.).
K.M. was supported by the Program of Excellence in Photon Science. We thank Shunsuke Neda for pointing out the errors in Eqs.~\eqref{eq : density contrast in nucleon bubble model} and ~\eqref{Tdecay as function of parameters}.
\end{acknowledgments}

\small
\bibliographystyle{JHEP}
\bibliography{Ref}

\providecommand{\href}[2]{#2}\begingroup\raggedright\begin{thebibliography}{100}

\bibitem{LIGOScientific:2016aoc}
{\scshape LIGO Scientific, Virgo} collaboration, B.~P. Abbott et~al.,
  \emph{{Observation of Gravitational Waves from a Binary Black Hole Merger}},
  \href{https://doi.org/10.1103/PhysRevLett.116.061102}{\emph{Phys. Rev. Lett.}
  {\bfseries 116} (2016) 061102},
  [\href{https://arxiv.org/abs/1602.03837}{{\ttfamily 1602.03837}}].

\bibitem{LIGOScientific:2018mvr}
{\scshape LIGO Scientific, Virgo} collaboration, B.~P. Abbott et~al.,
  \emph{{GWTC-1: A Gravitational-Wave Transient Catalog of Compact Binary
  Mergers Observed by LIGO and Virgo during the First and Second Observing
  Runs}}, \href{https://doi.org/10.1103/PhysRevX.9.031040}{\emph{Phys. Rev. X}
  {\bfseries 9} (2019) 031040},
  [\href{https://arxiv.org/abs/1811.12907}{{\ttfamily 1811.12907}}].

\bibitem{LIGOScientific:2020ibl}
{\scshape LIGO Scientific, Virgo} collaboration, R.~Abbott et~al.,
  \emph{{GWTC-2: Compact Binary Coalescences Observed by LIGO and Virgo During
  the First Half of the Third Observing Run}},
  \href{https://doi.org/10.1103/PhysRevX.11.021053}{\emph{Phys. Rev. X}
  {\bfseries 11} (2021) 021053},
  [\href{https://arxiv.org/abs/2010.14527}{{\ttfamily 2010.14527}}].

\bibitem{LIGOScientific:2021usb}
{\scshape LIGO Scientific, VIRGO} collaboration, R.~Abbott et~al.,
  \emph{{GWTC-2.1: Deep Extended Catalog of Compact Binary Coalescences
  Observed by LIGO and Virgo During the First Half of the Third Observing
  Run}},  \href{https://arxiv.org/abs/2108.01045}{{\ttfamily 2108.01045}}.

\bibitem{LIGOScientific:2021djp}
{\scshape LIGO Scientific, VIRGO, KAGRA} collaboration, R.~Abbott et~al.,
  \emph{{GWTC-3: Compact Binary Coalescences Observed by LIGO and Virgo During
  the Second Part of the Third Observing Run}},
  \href{https://arxiv.org/abs/2111.03606}{{\ttfamily 2111.03606}}.

\bibitem{Kormendy:1995er}
J.~Kormendy and D.~Richstone, \emph{{Inward bound: The Search for supermassive
  black holes in galactic nuclei}},
  \href{https://doi.org/10.1146/annurev.aa.33.090195.003053}{\emph{Ann. Rev.
  Astron. Astrophys.} {\bfseries 33} (1995) 581}.

\bibitem{Magorrian:1997hw}
J.~Magorrian et~al., \emph{{The Demography of massive dark objects in galaxy
  centers}}, \href{https://doi.org/10.1086/300353}{\emph{Astron. J.} {\bfseries
  115} (1998) 2285}, [\href{https://arxiv.org/abs/astro-ph/9708072}{{\ttfamily
  astro-ph/9708072}}].

\bibitem{Richstone:1998ky}
D.~Richstone et~al., \emph{{Supermassive black holes and the evolution of
  galaxies}}, {\emph{Nature} {\bfseries 395} (1998) A14--A19},
  [\href{https://arxiv.org/abs/astro-ph/9810378}{{\ttfamily
  astro-ph/9810378}}].

\bibitem{Matsuoka:2016pho}
Y.~Matsuoka et~al., \emph{{Subaru high-z exploration of low-luminosity quasars
  (SHELLQs). I. Discovery of 15 quasars and bright galaxies at 5.7
  \ensuremath{<} z \ensuremath{<} 6.9}},
  \href{https://doi.org/10.3847/0004-637X/828/1/26}{\emph{Astrophys. J.}
  {\bfseries 828} (2016) 26},
  [\href{https://arxiv.org/abs/1603.02281}{{\ttfamily 1603.02281}}].

\bibitem{Banados:2017unc}
E.~Banados et~al., \emph{{An 800-million-solar-mass black hole in a
  significantly neutral Universe at redshift 7.5}},
  \href{https://doi.org/10.1038/nature25180}{\emph{Nature} {\bfseries 553}
  (2018) 473--476}, [\href{https://arxiv.org/abs/1712.01860}{{\ttfamily
  1712.01860}}].

\bibitem{Volonteri:2010wz}
M.~Volonteri, \emph{{Formation of Supermassive Black Holes}},
  \href{https://doi.org/10.1007/s00159-010-0029-x}{\emph{Astron. Astrophys.
  Rev.} {\bfseries 18} (2010) 279--315},
  [\href{https://arxiv.org/abs/1003.4404}{{\ttfamily 1003.4404}}].

\bibitem{Bird:2016dcv}
S.~Bird, I.~Cholis, J.~B. Mu\~noz, Y.~Ali-Ha\"\i{}moud, M.~Kamionkowski, E.~D.
  Kovetz et~al., \emph{{Did LIGO detect dark matter?}},
  \href{https://doi.org/10.1103/PhysRevLett.116.201301}{\emph{Phys. Rev. Lett.}
  {\bfseries 116} (2016) 201301},
  [\href{https://arxiv.org/abs/1603.00464}{{\ttfamily 1603.00464}}].

\bibitem{Kashlinsky:2016sdv}
A.~Kashlinsky, \emph{{LIGO gravitational wave detection, primordial black holes
  and the near-IR cosmic infrared background anisotropies}},
  \href{https://doi.org/10.3847/2041-8205/823/2/L25}{\emph{Astrophys. J. Lett.}
  {\bfseries 823} (2016) L25},
  [\href{https://arxiv.org/abs/1605.04023}{{\ttfamily 1605.04023}}].

\bibitem{Sasaki:2016jop}
M.~Sasaki, T.~Suyama, T.~Tanaka and S.~Yokoyama, \emph{{Primordial Black Hole
  Scenario for the Gravitational-Wave Event GW150914}},
  \href{https://doi.org/10.1103/PhysRevLett.117.061101}{\emph{Phys. Rev. Lett.}
  {\bfseries 117} (2016) 061101},
  [\href{https://arxiv.org/abs/1603.08338}{{\ttfamily 1603.08338}}].

\bibitem{Carr:2016drx}
B.~Carr, F.~Kuhnel and M.~Sandstad, \emph{{Primordial Black Holes as Dark
  Matter}}, \href{https://doi.org/10.1103/PhysRevD.94.083504}{\emph{Phys. Rev.
  D} {\bfseries 94} (2016) 083504},
  [\href{https://arxiv.org/abs/1607.06077}{{\ttfamily 1607.06077}}].

\bibitem{Clesse:2016vqa}
S.~Clesse and J.~Garc\'\i{}a-Bellido, \emph{{The clustering of massive
  Primordial Black Holes as Dark Matter: measuring their mass distribution with
  Advanced LIGO}},
  \href{https://doi.org/10.1016/j.dark.2016.10.002}{\emph{Phys. Dark Univ.}
  {\bfseries 15} (2017) 142--147},
  [\href{https://arxiv.org/abs/1603.05234}{{\ttfamily 1603.05234}}].

\bibitem{Eroshenko:2016hmn}
Y.~N. Eroshenko, \emph{{Gravitational waves from primordial black holes
  collisions in binary systems}},
  \href{https://doi.org/10.1088/1742-6596/1051/1/012010}{\emph{J. Phys. Conf.
  Ser.} {\bfseries 1051} (2018) 012010},
  [\href{https://arxiv.org/abs/1604.04932}{{\ttfamily 1604.04932}}].

\bibitem{Ali-Haimoud:2017rtz}
Y.~Ali-Ha\"\i{}moud, E.~D. Kovetz and M.~Kamionkowski, \emph{{Merger rate of
  primordial black-hole binaries}},
  \href{https://doi.org/10.1103/PhysRevD.96.123523}{\emph{Phys. Rev. D}
  {\bfseries 96} (2017) 123523},
  [\href{https://arxiv.org/abs/1709.06576}{{\ttfamily 1709.06576}}].

\bibitem{Liu:2018ess}
L.~Liu, Z.-K. Guo and R.-G. Cai, \emph{{Effects of the surrounding primordial
  black holes on the merger rate of primordial black hole binaries}},
  \href{https://doi.org/10.1103/PhysRevD.99.063523}{\emph{Phys. Rev. D}
  {\bfseries 99} (2019) 063523},
  [\href{https://arxiv.org/abs/1812.05376}{{\ttfamily 1812.05376}}].

\bibitem{Vaskonen:2019jpv}
V.~Vaskonen and H.~Veerm\"ae, \emph{{Lower bound on the primordial black hole
  merger rate}}, \href{https://doi.org/10.1103/PhysRevD.101.043015}{\emph{Phys.
  Rev. D} {\bfseries 101} (2020) 043015},
  [\href{https://arxiv.org/abs/1908.09752}{{\ttfamily 1908.09752}}].

\bibitem{Liu:2019rnx}
L.~Liu, Z.-K. Guo and R.-G. Cai, \emph{{Effects of the merger history on the
  merger rate density of primordial black hole binaries}},
  \href{https://doi.org/10.1140/epjc/s10052-019-7227-0}{\emph{Eur. Phys. J. C}
  {\bfseries 79} (2019) 717},
  [\href{https://arxiv.org/abs/1901.07672}{{\ttfamily 1901.07672}}].

\bibitem{Wu:2020drm}
Y.~Wu, \emph{{Merger history of primordial black-hole binaries}},
  \href{https://doi.org/10.1103/PhysRevD.101.083008}{\emph{Phys. Rev. D}
  {\bfseries 101} (2020) 083008},
  [\href{https://arxiv.org/abs/2001.03833}{{\ttfamily 2001.03833}}].

\bibitem{DeLuca:2020bjf}
V.~De~Luca, G.~Franciolini, P.~Pani and A.~Riotto, \emph{{The evolution of
  primordial black holes and their final observable spins}},
  \href{https://doi.org/10.1088/1475-7516/2020/04/052}{\emph{JCAP} {\bfseries
  04} (2020) 052}, [\href{https://arxiv.org/abs/2003.02778}{{\ttfamily
  2003.02778}}].

\bibitem{DeLuca:2020qqa}
V.~De~Luca, G.~Franciolini, P.~Pani and A.~Riotto, \emph{{Primordial Black
  Holes Confront LIGO/Virgo data: Current situation}},
  \href{https://doi.org/10.1088/1475-7516/2020/06/044}{\emph{JCAP} {\bfseries
  06} (2020) 044}, [\href{https://arxiv.org/abs/2005.05641}{{\ttfamily
  2005.05641}}].

\bibitem{Hawking:1971ei}
S.~Hawking, \emph{{Gravitationally collapsed objects of very low mass}},
  {\emph{Mon. Not. Roy. Astron. Soc.} {\bfseries 152} (1971) 75}.

\bibitem{Carr:1974nx}
B.~J. Carr and S.~W. Hawking, \emph{{Black holes in the early Universe}},
  {\emph{Mon. Not. Roy. Astron. Soc.} {\bfseries 168} (1974) 399--415}.

\bibitem{Carr:1975qj}
B.~J. Carr, \emph{{The Primordial black hole mass spectrum}},
  \href{https://doi.org/10.1086/153853}{\emph{Astrophys. J.} {\bfseries 201}
  (1975) 1--19}.

\bibitem{Garcia-Bellido:1996mdl}
J.~Garcia-Bellido, A.~D. Linde and D.~Wands, \emph{{Density perturbations and
  black hole formation in hybrid inflation}},
  \href{https://doi.org/10.1103/PhysRevD.54.6040}{\emph{Phys. Rev. D}
  {\bfseries 54} (1996) 6040--6058},
  [\href{https://arxiv.org/abs/astro-ph/9605094}{{\ttfamily
  astro-ph/9605094}}].

\bibitem{Yokoyama:1995ex}
J.~Yokoyama, \emph{{Formation of MACHO primordial black holes in inflationary
  cosmology}}, {\emph{Astron. Astrophys.} {\bfseries 318} (1997) 673},
  [\href{https://arxiv.org/abs/astro-ph/9509027}{{\ttfamily
  astro-ph/9509027}}].

\bibitem{Kawasaki:1997ju}
M.~Kawasaki, N.~Sugiyama and T.~Yanagida, \emph{{Primordial black hole
  formation in a double inflation model in supergravity}},
  \href{https://doi.org/10.1103/PhysRevD.57.6050}{\emph{Phys. Rev. D}
  {\bfseries 57} (1998) 6050--6056},
  [\href{https://arxiv.org/abs/hep-ph/9710259}{{\ttfamily hep-ph/9710259}}].

\bibitem{Kawasaki:2016pql}
M.~Kawasaki, A.~Kusenko, Y.~Tada and T.~T. Yanagida, \emph{{Primordial black
  holes as dark matter in supergravity inflation models}},
  \href{https://doi.org/10.1103/PhysRevD.94.083523}{\emph{Phys. Rev. D}
  {\bfseries 94} (2016) 083523},
  [\href{https://arxiv.org/abs/1606.07631}{{\ttfamily 1606.07631}}].

\bibitem{Inomata:2016rbd}
K.~Inomata, M.~Kawasaki, K.~Mukaida, Y.~Tada and T.~T. Yanagida,
  \emph{{Inflationary primordial black holes for the LIGO gravitational wave
  events and pulsar timing array experiments}},
  \href{https://doi.org/10.1103/PhysRevD.95.123510}{\emph{Phys. Rev. D}
  {\bfseries 95} (2017) 123510},
  [\href{https://arxiv.org/abs/1611.06130}{{\ttfamily 1611.06130}}].

\bibitem{Inomata:2018cht}
K.~Inomata, M.~Kawasaki, K.~Mukaida and T.~T. Yanagida, \emph{{Double inflation
  as a single origin of primordial black holes for all dark matter and LIGO
  observations}}, \href{https://doi.org/10.1103/PhysRevD.97.043514}{\emph{Phys.
  Rev. D} {\bfseries 97} (2018) 043514},
  [\href{https://arxiv.org/abs/1711.06129}{{\ttfamily 1711.06129}}].

\bibitem{Sasaki:2018dmp}
M.~Sasaki, T.~Suyama, T.~Tanaka and S.~Yokoyama, \emph{{Primordial black
  holes\textemdash{}perspectives in gravitational wave astronomy}},
  \href{https://doi.org/10.1088/1361-6382/aaa7b4}{\emph{Class. Quant. Grav.}
  {\bfseries 35} (2018) 063001},
  [\href{https://arxiv.org/abs/1801.05235}{{\ttfamily 1801.05235}}].

\bibitem{Ando:2017veq}
K.~Ando, K.~Inomata, M.~Kawasaki, K.~Mukaida and T.~T. Yanagida,
  \emph{{Primordial black holes for the LIGO events in the axionlike curvaton
  model}}, \href{https://doi.org/10.1103/PhysRevD.97.123512}{\emph{Phys. Rev.
  D} {\bfseries 97} (2018) 123512},
  [\href{https://arxiv.org/abs/1711.08956}{{\ttfamily 1711.08956}}].

\bibitem{Ando:2018nge}
K.~Ando, M.~Kawasaki and H.~Nakatsuka, \emph{{Formation of primordial black
  holes in an axionlike curvaton model}},
  \href{https://doi.org/10.1103/PhysRevD.98.083508}{\emph{Phys. Rev. D}
  {\bfseries 98} (2018) 083508},
  [\href{https://arxiv.org/abs/1805.07757}{{\ttfamily 1805.07757}}].

\bibitem{Kohri:2012yw}
K.~Kohri, C.-M. Lin and T.~Matsuda, \emph{{Primordial black holes from the
  inflating curvaton}},
  \href{https://doi.org/10.1103/PhysRevD.87.103527}{\emph{Phys. Rev. D}
  {\bfseries 87} (2013) 103527},
  [\href{https://arxiv.org/abs/1211.2371}{{\ttfamily 1211.2371}}].

\bibitem{Chen:2019zza}
C.~Chen and Y.-F. Cai, \emph{{Primordial black holes from sound speed resonance
  in the inflaton-curvaton mixed scenario}},
  \href{https://doi.org/10.1088/1475-7516/2019/10/068}{\emph{JCAP} {\bfseries
  10} (2019) 068}, [\href{https://arxiv.org/abs/1908.03942}{{\ttfamily
  1908.03942}}].

\bibitem{Liu:2021rgq}
L.-H. Liu and W.-L. Xu, \emph{{The primordial black hole from running
  curvaton}},  \href{https://arxiv.org/abs/2107.07310}{{\ttfamily 2107.07310}}.

\bibitem{Pi:2021dft}
S.~Pi and M.~Sasaki, \emph{{Primordial Black Hole Formation in Non-Minimal
  Curvaton Scenario}},  \href{https://arxiv.org/abs/2112.12680}{{\ttfamily
  2112.12680}}.

\bibitem{Suyama:2011pu}
T.~Suyama and J.~Yokoyama, \emph{{Temporal enhancement of super-horizon
  curvature perturbations from decays of two curvatons and its cosmological
  consequences}}, \href{https://doi.org/10.1103/PhysRevD.84.083511}{\emph{Phys.
  Rev. D} {\bfseries 84} (2011) 083511},
  [\href{https://arxiv.org/abs/1106.5983}{{\ttfamily 1106.5983}}].

\bibitem{Maeso:2021xvl}
D.~N. Maeso, L.~Marzola, M.~Raidal, V.~Vaskonen and H.~Veerm\"ae,
  \emph{{Primordial black holes from spectator field bubbles}},
  \href{https://doi.org/10.1088/1475-7516/2022/02/017}{\emph{JCAP} {\bfseries
  02} (2022) 017}, [\href{https://arxiv.org/abs/2112.01505}{{\ttfamily
  2112.01505}}].

\bibitem{Cai:2021wzd}
R.-G. Cai, C.~Chen and C.~Fu, \emph{{Primordial black holes and stochastic
  gravitational wave background from inflation with a noncanonical spectator
  field}}, \href{https://doi.org/10.1103/PhysRevD.104.083537}{\emph{Phys. Rev.
  D} {\bfseries 104} (2021) 083537},
  [\href{https://arxiv.org/abs/2108.03422}{{\ttfamily 2108.03422}}].

\bibitem{Carr:2017edp}
B.~Carr, T.~Tenkanen and V.~Vaskonen, \emph{{Primordial black holes from
  inflaton and spectator field perturbations in a matter-dominated era}},
  \href{https://doi.org/10.1103/PhysRevD.96.063507}{\emph{Phys. Rev. D}
  {\bfseries 96} (2017) 063507},
  [\href{https://arxiv.org/abs/1706.03746}{{\ttfamily 1706.03746}}].

\bibitem{Carr:2020gox}
B.~Carr, K.~Kohri, Y.~Sendouda and J.~Yokoyama, \emph{{Constraints on
  primordial black holes}},
  \href{https://doi.org/10.1088/1361-6633/ac1e31}{\emph{Rept. Prog. Phys.}
  {\bfseries 84} (2021) 116902},
  [\href{https://arxiv.org/abs/2002.12778}{{\ttfamily 2002.12778}}].

\bibitem{Chluba:2012we}
J.~Chluba, A.~L. Erickcek and I.~Ben-Dayan, \emph{{Probing the inflaton:
  Small-scale power spectrum constraints from measurements of the CMB energy
  spectrum}},
  \href{https://doi.org/10.1088/0004-637X/758/2/76}{\emph{Astrophys. J.}
  {\bfseries 758} (2012) 76},
  [\href{https://arxiv.org/abs/1203.2681}{{\ttfamily 1203.2681}}].

\bibitem{Kohri:2014lza}
K.~Kohri, T.~Nakama and T.~Suyama, \emph{{Testing scenarios of primordial black
  holes being the seeds of supermassive black holes by ultracompact minihalos
  and CMB $\mu$-distortions}},
  \href{https://doi.org/10.1103/PhysRevD.90.083514}{\emph{Phys. Rev. D}
  {\bfseries 90} (2014) 083514},
  [\href{https://arxiv.org/abs/1405.5999}{{\ttfamily 1405.5999}}].

\bibitem{Saito:2008jc}
R.~Saito and J.~Yokoyama, \emph{{Gravitational wave background as a probe of
  the primordial black hole abundance}},
  \href{https://doi.org/10.1103/PhysRevLett.102.161101}{\emph{Phys. Rev. Lett.}
  {\bfseries 102} (2009) 161101},
  [\href{https://arxiv.org/abs/0812.4339}{{\ttfamily 0812.4339}}].

\bibitem{Bugaev:2010bb}
E.~Bugaev and P.~Klimai, \emph{{Constraints on the induced gravitational wave
  background from primordial black holes}},
  \href{https://doi.org/10.1103/PhysRevD.83.083521}{\emph{Phys. Rev. D}
  {\bfseries 83} (2011) 083521},
  [\href{https://arxiv.org/abs/1012.4697}{{\ttfamily 1012.4697}}].

\bibitem{Lentati:2015qwp}
L.~Lentati et~al., \emph{{European Pulsar Timing Array Limits On An Isotropic
  Stochastic Gravitational-Wave Background}},
  \href{https://doi.org/10.1093/mnras/stv1538}{\emph{Mon. Not. Roy. Astron.
  Soc.} {\bfseries 453} (2015) 2576--2598},
  [\href{https://arxiv.org/abs/1504.03692}{{\ttfamily 1504.03692}}].

\bibitem{NANOGrav:2015aud}
{\scshape NANOGrav} collaboration, Z.~Arzoumanian et~al., \emph{{The NANOGrav
  Nine-year Data Set: Limits on the Isotropic Stochastic Gravitational Wave
  Background}},
  \href{https://doi.org/10.3847/0004-637X/821/1/13}{\emph{Astrophys. J.}
  {\bfseries 821} (2016) 13},
  [\href{https://arxiv.org/abs/1508.03024}{{\ttfamily 1508.03024}}].

\bibitem{Shannon:2015ect}
R.~M. Shannon et~al., \emph{{Gravitational waves from binary supermassive black
  holes missing in pulsar observations}},
  \href{https://doi.org/10.1126/science.aab1910}{\emph{Science} {\bfseries 349}
  (2015) 1522--1525}, [\href{https://arxiv.org/abs/1509.07320}{{\ttfamily
  1509.07320}}].

\bibitem{Lu:2020bmd}
P.~Lu, V.~Takhistov, G.~B. Gelmini, K.~Hayashi, Y.~Inoue and A.~Kusenko,
  \emph{{Constraining Primordial Black Holes with Dwarf Galaxy Heating}},
  \href{https://doi.org/10.3847/2041-8213/abdcb6}{\emph{Astrophys. J. Lett.}
  {\bfseries 908} (2021) L23},
  [\href{https://arxiv.org/abs/2007.02213}{{\ttfamily 2007.02213}}].

\bibitem{Oguri:2017ock}
M.~Oguri, J.~M. Diego, N.~Kaiser, P.~L. Kelly and T.~Broadhurst,
  \emph{{Understanding caustic crossings in giant arcs: characteristic scales,
  event rates, and constraints on compact dark matter}},
  \href{https://doi.org/10.1103/PhysRevD.97.023518}{\emph{Phys. Rev. D}
  {\bfseries 97} (2018) 023518},
  [\href{https://arxiv.org/abs/1710.00148}{{\ttfamily 1710.00148}}].

\bibitem{Ali-Haimoud:2016mbv}
Y.~Ali-Ha\"\i{}moud and M.~Kamionkowski, \emph{{Cosmic microwave background
  limits on accreting primordial black holes}},
  \href{https://doi.org/10.1103/PhysRevD.95.043534}{\emph{Phys. Rev. D}
  {\bfseries 95} (2017) 043534},
  [\href{https://arxiv.org/abs/1612.05644}{{\ttfamily 1612.05644}}].

\bibitem{Murgia:2019duy}
R.~Murgia, G.~Scelfo, M.~Viel and A.~Raccanelli,
  \emph{{Lyman-\ensuremath{\alpha} Forest Constraints on Primordial Black Holes
  as Dark Matter}},
  \href{https://doi.org/10.1103/PhysRevLett.123.071102}{\emph{Phys. Rev. Lett.}
  {\bfseries 123} (2019) 071102},
  [\href{https://arxiv.org/abs/1903.10509}{{\ttfamily 1903.10509}}].

\bibitem{Brandt:2016aco}
T.~D. Brandt, \emph{{Constraints on MACHO Dark Matter from Compact Stellar
  Systems in Ultra-Faint Dwarf Galaxies}},
  \href{https://doi.org/10.3847/2041-8205/824/2/L31}{\emph{Astrophys. J. Lett.}
  {\bfseries 824} (2016) L31},
  [\href{https://arxiv.org/abs/1605.03665}{{\ttfamily 1605.03665}}].

\bibitem{Koushiappas:2017chw}
S.~M. Koushiappas and A.~Loeb, \emph{{Dynamics of Dwarf Galaxies Disfavor
  Stellar-Mass Black Holes as Dark Matter}},
  \href{https://doi.org/10.1103/PhysRevLett.119.041102}{\emph{Phys. Rev. Lett.}
  {\bfseries 119} (2017) 041102},
  [\href{https://arxiv.org/abs/1704.01668}{{\ttfamily 1704.01668}}].

\bibitem{Monroy-Rodriguez:2014ula}
M.~A. Monroy-Rodr\'\i{}guez and C.~Allen, \emph{{The end of the MACHO era-
  revisited: new limits on MACHO masses from halo wide binaries}},
  \href{https://doi.org/10.1088/0004-637X/790/2/159}{\emph{Astrophys. J.}
  {\bfseries 790} (2014) 159},
  [\href{https://arxiv.org/abs/1406.5169}{{\ttfamily 1406.5169}}].

\bibitem{Smith:2016hsc}
B.~D. Smith et~al., \emph{{Grackle: a Chemistry and Cooling Library for
  Astrophysics}}, \href{https://doi.org/10.1093/mnras/stw3291}{\emph{Mon. Not.
  Roy. Astron. Soc.} {\bfseries 466} (2017) 2217--2234},
  [\href{https://arxiv.org/abs/1610.09591}{{\ttfamily 1610.09591}}].

\bibitem{Carr:2018rid}
B.~Carr and J.~Silk, \emph{{Primordial Black Holes as Generators of Cosmic
  Structures}}, \href{https://doi.org/10.1093/mnras/sty1204}{\emph{Mon. Not.
  Roy. Astron. Soc.} {\bfseries 478} (2018) 3756--3775},
  [\href{https://arxiv.org/abs/1801.00672}{{\ttfamily 1801.00672}}].

\bibitem{Ricotti:2007au}
M.~Ricotti, J.~P. Ostriker and K.~J. Mack, \emph{{Effect of Primordial Black
  Holes on the Cosmic Microwave Background and Cosmological Parameter
  Estimates}}, \href{https://doi.org/10.1086/587831}{\emph{Astrophys. J.}
  {\bfseries 680} (2008) 829},
  [\href{https://arxiv.org/abs/0709.0524}{{\ttfamily 0709.0524}}].

\bibitem{Garcia-Bellido:2017aan}
J.~Garcia-Bellido, M.~Peloso and C.~Unal, \emph{{Gravitational Wave signatures
  of inflationary models from Primordial Black Hole Dark Matter}},
  \href{https://doi.org/10.1088/1475-7516/2017/09/013}{\emph{JCAP} {\bfseries
  09} (2017) 013}, [\href{https://arxiv.org/abs/1707.02441}{{\ttfamily
  1707.02441}}].

\bibitem{Hawking:1987bn}
S.~W. Hawking, \emph{{Black Holes From Cosmic Strings}},
  \href{https://doi.org/10.1016/0370-2693(89)90206-2}{\emph{Phys. Lett. B}
  {\bfseries 231} (1989) 237--239}.

\bibitem{Caldwell:1995fu}
R.~R. Caldwell and P.~Casper, \emph{{Formation of black holes from collapsed
  cosmic string loops}},
  \href{https://doi.org/10.1103/PhysRevD.53.3002}{\emph{Phys. Rev. D}
  {\bfseries 53} (1996) 3002--3010},
  [\href{https://arxiv.org/abs/gr-qc/9509012}{{\ttfamily gr-qc/9509012}}].

\bibitem{Garriga:1992nm}
J.~Garriga and A.~Vilenkin, \emph{{Black holes from nucleating strings}},
  \href{https://doi.org/10.1103/PhysRevD.47.3265}{\emph{Phys. Rev. D}
  {\bfseries 47} (1993) 3265--3274},
  [\href{https://arxiv.org/abs/hep-ph/9208212}{{\ttfamily hep-ph/9208212}}].

\bibitem{Cotner:2016cvr}
E.~Cotner and A.~Kusenko, \emph{{Primordial black holes from supersymmetry in
  the early universe}},
  \href{https://doi.org/10.1103/PhysRevLett.119.031103}{\emph{Phys. Rev. Lett.}
  {\bfseries 119} (2017) 031103},
  [\href{https://arxiv.org/abs/1612.02529}{{\ttfamily 1612.02529}}].

\bibitem{Nakama:2016kfq}
T.~Nakama, T.~Suyama and J.~Yokoyama, \emph{{Supermassive black holes formed by
  direct collapse of inflationary perturbations}},
  \href{https://doi.org/10.1103/PhysRevD.94.103522}{\emph{Phys. Rev. D}
  {\bfseries 94} (2016) 103522},
  [\href{https://arxiv.org/abs/1609.02245}{{\ttfamily 1609.02245}}].

\bibitem{Deng:2017uwc}
H.~Deng and A.~Vilenkin, \emph{{Primordial black hole formation by vacuum
  bubbles}}, \href{https://doi.org/10.1088/1475-7516/2017/12/044}{\emph{JCAP}
  {\bfseries 12} (2017) 044},
  [\href{https://arxiv.org/abs/1710.02865}{{\ttfamily 1710.02865}}].

\bibitem{Kitajima:2020kig}
N.~Kitajima and F.~Takahashi, \emph{{Primordial Black Holes from QCD Axion
  Bubbles}}, \href{https://doi.org/10.1088/1475-7516/2020/11/060}{\emph{JCAP}
  {\bfseries 11} (2020) 060},
  [\href{https://arxiv.org/abs/2006.13137}{{\ttfamily 2006.13137}}].

\bibitem{Kawana:2021tde}
K.~Kawana and K.-P. Xie, \emph{{Primordial black holes from a cosmic phase
  transition: The collapse of Fermi-balls}},
  \href{https://doi.org/10.1016/j.physletb.2021.136791}{\emph{Phys. Lett. B}
  {\bfseries 824} (2022) 136791},
  [\href{https://arxiv.org/abs/2106.00111}{{\ttfamily 2106.00111}}].

\bibitem{Huang:2022him}
P.~Huang and K.-P. Xie, \emph{{Primordial black holes from an electroweak phase
  transition}},  \href{https://arxiv.org/abs/2201.07243}{{\ttfamily
  2201.07243}}.

\bibitem{Cotner:2018vug}
E.~Cotner, A.~Kusenko and V.~Takhistov, \emph{{Primordial Black Holes from
  Inflaton Fragmentation into Oscillons}},
  \href{https://doi.org/10.1103/PhysRevD.98.083513}{\emph{Phys. Rev. D}
  {\bfseries 98} (2018) 083513},
  [\href{https://arxiv.org/abs/1801.03321}{{\ttfamily 1801.03321}}].

\bibitem{Cotner:2019ykd}
E.~Cotner, A.~Kusenko, M.~Sasaki and V.~Takhistov, \emph{{Analytic Description
  of Primordial Black Hole Formation from Scalar Field Fragmentation}},
  \href{https://doi.org/10.1088/1475-7516/2019/10/077}{\emph{JCAP} {\bfseries
  10} (2019) 077}, [\href{https://arxiv.org/abs/1907.10613}{{\ttfamily
  1907.10613}}].

\bibitem{Kusenko:2020pcg}
A.~Kusenko, M.~Sasaki, S.~Sugiyama, M.~Takada, V.~Takhistov and E.~Vitagliano,
  \emph{{Exploring Primordial Black Holes from the Multiverse with Optical
  Telescopes}},
  \href{https://doi.org/10.1103/PhysRevLett.125.181304}{\emph{Phys. Rev. Lett.}
  {\bfseries 125} (2020) 181304},
  [\href{https://arxiv.org/abs/2001.09160}{{\ttfamily 2001.09160}}].

\bibitem{Dolgov:1992pu}
A.~Dolgov and J.~Silk, \emph{{Baryon isocurvature fluctuations at small scales
  and baryonic dark matter}},
  \href{https://doi.org/10.1103/PhysRevD.47.4244}{\emph{Phys. Rev. D}
  {\bfseries 47} (1993) 4244--4255}.

\bibitem{Dolgov:2008wu}
A.~D. Dolgov, M.~Kawasaki and N.~Kevlishvili, \emph{{Inhomogeneous
  baryogenesis, cosmic antimatter, and dark matter}},
  \href{https://doi.org/10.1016/j.nuclphysb.2008.08.029}{\emph{Nucl. Phys. B}
  {\bfseries 807} (2009) 229--250},
  [\href{https://arxiv.org/abs/0806.2986}{{\ttfamily 0806.2986}}].

\bibitem{Hasegawa:2017jtk}
F.~Hasegawa and M.~Kawasaki, \emph{{Cogenesis of LIGO Primordial Black Holes
  and Dark Matter}},
  \href{https://doi.org/10.1103/PhysRevD.98.043514}{\emph{Phys. Rev. D}
  {\bfseries 98} (2018) 043514},
  [\href{https://arxiv.org/abs/1711.00990}{{\ttfamily 1711.00990}}].

\bibitem{Hasegawa:2018yuy}
F.~Hasegawa and M.~Kawasaki, \emph{{Primordial Black Holes from Affleck-Dine
  Mechanism}}, \href{https://doi.org/10.1088/1475-7516/2019/01/027}{\emph{JCAP}
  {\bfseries 01} (2019) 027},
  [\href{https://arxiv.org/abs/1807.00463}{{\ttfamily 1807.00463}}].

\bibitem{Kawasaki:2019iis}
M.~Kawasaki and K.~Murai, \emph{{Formation of supermassive primordial black
  holes by Affleck-Dine mechanism}},
  \href{https://doi.org/10.1103/PhysRevD.100.103521}{\emph{Phys. Rev. D}
  {\bfseries 100} (2019) 103521},
  [\href{https://arxiv.org/abs/1907.02273}{{\ttfamily 1907.02273}}].

\bibitem{Kawasaki:2021zir}
M.~Kawasaki, K.~Murai and H.~Nakatsuka, \emph{{Strong clustering of primordial
  black holes from Affleck-Dine mechanism}},
  \href{https://doi.org/10.1088/1475-7516/2021/10/025}{\emph{JCAP} {\bfseries
  10} (2021) 025}, [\href{https://arxiv.org/abs/2107.03580}{{\ttfamily
  2107.03580}}].

\bibitem{Kawasaki:2002hq}
M.~Kawasaki, F.~Takahashi and M.~Yamaguchi, \emph{{Large lepton asymmetry from
  Q balls}}, \href{https://doi.org/10.1103/PhysRevD.66.043516}{\emph{Phys. Rev.
  D} {\bfseries 66} (2002) 043516},
  [\href{https://arxiv.org/abs/hep-ph/0205101}{{\ttfamily hep-ph/0205101}}].

\bibitem{Coleman:1985ki}
S.~R. Coleman, \emph{{Q-balls}},
  \href{https://doi.org/10.1016/0550-3213(86)90520-1}{\emph{Nucl. Phys. B}
  {\bfseries 262} (1985) 263}.

\bibitem{Kusenko:1997zq}
A.~Kusenko, \emph{{Solitons in the supersymmetric extensions of the standard
  model}}, \href{https://doi.org/10.1016/S0370-2693(97)00584-4}{\emph{Phys.
  Lett. B} {\bfseries 405} (1997) 108},
  [\href{https://arxiv.org/abs/hep-ph/9704273}{{\ttfamily hep-ph/9704273}}].

\bibitem{Kusenko:1997si}
A.~Kusenko and M.~E. Shaposhnikov, \emph{{Supersymmetric Q balls as dark
  matter}}, \href{https://doi.org/10.1016/S0370-2693(97)01375-0}{\emph{Phys.
  Lett. B} {\bfseries 418} (1998) 46--54},
  [\href{https://arxiv.org/abs/hep-ph/9709492}{{\ttfamily hep-ph/9709492}}].

\bibitem{Enqvist:1997si}
K.~Enqvist and J.~McDonald, \emph{{Q balls and baryogenesis in the MSSM}},
  \href{https://doi.org/10.1016/S0370-2693(98)00271-8}{\emph{Phys. Lett. B}
  {\bfseries 425} (1998) 309--321},
  [\href{https://arxiv.org/abs/hep-ph/9711514}{{\ttfamily hep-ph/9711514}}].

\bibitem{Kasuya:1999wu}
S.~Kasuya and M.~Kawasaki, \emph{{Q ball formation through Affleck-Dine
  mechanism}}, \href{https://doi.org/10.1103/PhysRevD.61.041301}{\emph{Phys.
  Rev. D} {\bfseries 61} (2000) 041301},
  [\href{https://arxiv.org/abs/hep-ph/9909509}{{\ttfamily hep-ph/9909509}}].

\bibitem{Laine:1998rg}
M.~Laine and M.~E. Shaposhnikov, \emph{{Thermodynamics of nontopological
  solitons}}, \href{https://doi.org/10.1016/S0550-3213(98)00474-X}{\emph{Nucl.
  Phys. B} {\bfseries 532} (1998) 376--404},
  [\href{https://arxiv.org/abs/hep-ph/9804237}{{\ttfamily hep-ph/9804237}}].

\bibitem{Cohen:1986ct}
A.~G. Cohen, S.~R. Coleman, H.~Georgi and A.~Manohar, \emph{{The Evaporation of
  $Q$ Balls}}, \href{https://doi.org/10.1016/0550-3213(86)90004-0}{\emph{Nucl.
  Phys. B} {\bfseries 272} (1986) 301--321}.

\bibitem{Kawasaki:2012gk}
M.~Kawasaki and M.~Yamada, \emph{{$Q$ ball Decay Rates into Gravitinos and
  Quarks}}, \href{https://doi.org/10.1103/PhysRevD.87.023517}{\emph{Phys. Rev.
  D} {\bfseries 87} (2013) 023517},
  [\href{https://arxiv.org/abs/1209.5781}{{\ttfamily 1209.5781}}].

\bibitem{Affleck:1984fy}
I.~Affleck and M.~Dine, \emph{{A New Mechanism for Baryogenesis}},
  \href{https://doi.org/10.1016/0550-3213(85)90021-5}{\emph{Nucl. Phys. B}
  {\bfseries 249} (1985) 361--380}.

\bibitem{Dine:1995kz}
M.~Dine, L.~Randall and S.~D. Thomas, \emph{{Baryogenesis from flat directions
  of the supersymmetric standard model}},
  \href{https://doi.org/10.1016/0550-3213(95)00538-2}{\emph{Nucl. Phys. B}
  {\bfseries 458} (1996) 291--326},
  [\href{https://arxiv.org/abs/hep-ph/9507453}{{\ttfamily hep-ph/9507453}}].

\bibitem{Harada:2013epa}
T.~Harada, C.-M. Yoo and K.~Kohri, \emph{{Threshold of primordial black hole
  formation}}, \href{https://doi.org/10.1103/PhysRevD.88.084051}{\emph{Phys.
  Rev. D} {\bfseries 88} (2013) 084051},
  [\href{https://arxiv.org/abs/1309.4201}{{\ttfamily 1309.4201}}].

\bibitem{Starobinsky:1982ee}
A.~A. Starobinsky, \emph{{Dynamics of Phase Transition in the New Inflationary
  Universe Scenario and Generation of Perturbations}},
  \href{https://doi.org/10.1016/0370-2693(82)90541-X}{\emph{Phys. Lett. B}
  {\bfseries 117} (1982) 175--178}.

\bibitem{Linde:1982uu}
A.~D. Linde, \emph{{Scalar Field Fluctuations in Expanding Universe and the New
  Inflationary Universe Scenario}},
  \href{https://doi.org/10.1016/0370-2693(82)90293-3}{\emph{Phys. Lett. B}
  {\bfseries 116} (1982) 335--339}.

\bibitem{Starobinsky:1994bd}
A.~A. Starobinsky and J.~Yokoyama, \emph{{Equilibrium state of a
  selfinteracting scalar field in the De Sitter background}},
  \href{https://doi.org/10.1103/PhysRevD.50.6357}{\emph{Phys. Rev. D}
  {\bfseries 50} (1994) 6357--6368},
  [\href{https://arxiv.org/abs/astro-ph/9407016}{{\ttfamily
  astro-ph/9407016}}].

\bibitem{Willott:2010yu}
C.~J. Willott, L.~Albert, D.~Arzoumanian, J.~Bergeron, D.~Crampton, P.~Delorme
  et~al., \emph{{Eddington-limited accretion and the black hole mass function
  at redshift 6}},
  \href{https://doi.org/10.1088/0004-6256/140/2/546}{\emph{Astron. J.}
  {\bfseries 140} (2010) 546},
  [\href{https://arxiv.org/abs/1006.1342}{{\ttfamily 1006.1342}}].

\bibitem{10.1093/mnras/stw1679}
Y.~Rosas-Guevara, R.~G. Bower, J.~Schaye, S.~McAlpine, C.~Dalla~Vecchia, C.~S.
  Frenk et~al., \emph{{Supermassive black holes in the EAGLE Universe.
  Revealing the observables of their growth}},
  \href{https://doi.org/10.1093/mnras/stw1679}{\emph{Monthly Notices of the
  Royal Astronomical Society} {\bfseries 462} (07, 2016) 190--205},
  [\href{https://arxiv.org/abs/https://academic.oup.com/mnras/article-pdf/462/1/190/18471698/stw1679.pdf}{{\ttfamily
  https://academic.oup.com/mnras/article-pdf/462/1/190/18471698/stw1679.pdf}}].

\bibitem{Serpico:2020ehh}
P.~D. Serpico, V.~Poulin, D.~Inman and K.~Kohri, \emph{{Cosmic microwave
  background bounds on primordial black holes including dark matter halo
  accretion}},
  \href{https://doi.org/10.1103/PhysRevResearch.2.023204}{\emph{Phys. Rev.
  Res.} {\bfseries 2} (2020) 023204},
  [\href{https://arxiv.org/abs/2002.10771}{{\ttfamily 2002.10771}}].

\bibitem{Applegate:1987hm}
J.~H. Applegate, C.~J. Hogan and R.~J. Scherrer, \emph{{Cosmological Baryon
  Diffusion and Nucleosynthesis}},
  \href{https://doi.org/10.1103/PhysRevD.35.1151}{\emph{Phys. Rev. D}
  {\bfseries 35} (1987) 1151--1160}.

\bibitem{Kasuya:2014ofa}
S.~Kasuya and M.~Kawasaki, \emph{{Baryogenesis from the gauge-mediation type
  Q-ball and the new type of Q-ball as the dark matter}},
  \href{https://doi.org/10.1103/PhysRevD.89.103534}{\emph{Phys. Rev. D}
  {\bfseries 89} (2014) 103534},
  [\href{https://arxiv.org/abs/1402.4546}{{\ttfamily 1402.4546}}].

\bibitem{Banerjee:2000mb}
R.~Banerjee and K.~Jedamzik, \emph{{On B-ball dark matter and baryogenesis}},
  \href{https://doi.org/10.1016/S0370-2693(00)00688-2}{\emph{Phys. Lett. B}
  {\bfseries 484} (2000) 278--282},
  [\href{https://arxiv.org/abs/hep-ph/0005031}{{\ttfamily hep-ph/0005031}}].

\bibitem{deGouvea:1997afu}
A.~de~Gouvea, T.~Moroi and H.~Murayama, \emph{{Cosmology of supersymmetric
  models with low-energy gauge mediation}},
  \href{https://doi.org/10.1103/PhysRevD.56.1281}{\emph{Phys. Rev. D}
  {\bfseries 56} (1997) 1281--1299},
  [\href{https://arxiv.org/abs/hep-ph/9701244}{{\ttfamily hep-ph/9701244}}].

\bibitem{Kasuya:2000sc}
S.~Kasuya and M.~Kawasaki, \emph{{A New type of stable Q balls in the gauge
  mediated SUSY breaking}},
  \href{https://doi.org/10.1103/PhysRevLett.85.2677}{\emph{Phys. Rev. Lett.}
  {\bfseries 85} (2000) 2677--2680},
  [\href{https://arxiv.org/abs/hep-ph/0006128}{{\ttfamily hep-ph/0006128}}].

\bibitem{Hiramatsu:2010dx}
T.~Hiramatsu, M.~Kawasaki and F.~Takahashi, \emph{{Numerical study of Q-ball
  formation in gravity mediation}},
  \href{https://doi.org/10.1088/1475-7516/2010/06/008}{\emph{JCAP} {\bfseries
  06} (2010) 008}, [\href{https://arxiv.org/abs/1003.1779}{{\ttfamily
  1003.1779}}].

\bibitem{Kasuya:2012mh}
S.~Kasuya, M.~Kawasaki and M.~Yamada, \emph{{Revisiting the gravitino dark
  matter and baryon asymmetry from Q-ball decay in gauge mediation}},
  \href{https://doi.org/10.1016/j.physletb.2013.08.008}{\emph{Phys. Lett. B}
  {\bfseries 726} (2013) 1--7},
  [\href{https://arxiv.org/abs/1211.4743}{{\ttfamily 1211.4743}}].

\bibitem{Planck:2018vyg}
{\scshape Planck} collaboration, N.~Aghanim et~al., \emph{{Planck 2018 results.
  VI. Cosmological parameters}},
  \href{https://doi.org/10.1051/0004-6361/201833910}{\emph{Astron. Astrophys.}
  {\bfseries 641} (2020) A6},
  [\href{https://arxiv.org/abs/1807.06209}{{\ttfamily 1807.06209}}].

\bibitem{Akita:2020szl}
K.~Akita and M.~Yamaguchi, \emph{{A precision calculation of relic neutrino
  decoupling}},
  \href{https://doi.org/10.1088/1475-7516/2020/08/012}{\emph{JCAP} {\bfseries
  08} (2020) 012}, [\href{https://arxiv.org/abs/2005.07047}{{\ttfamily
  2005.07047}}].

\bibitem{Moroi:1993mb}
T.~Moroi, H.~Murayama and M.~Yamaguchi, \emph{{Cosmological constraints on the
  light s table gravitino}},
  \href{https://doi.org/10.1016/0370-2693(93)91434-O}{\emph{Phys. Lett. B}
  {\bfseries 303} (1993) 289--294}.

\bibitem{Kawasaki:2022hvx}
M.~Kawasaki and K.~Murai, \emph{{Lepton Asymmetric Universe}},
  \href{https://arxiv.org/abs/2203.09713}{{\ttfamily 2203.09713}}.

\bibitem{Bolz:2000fu}
M.~Bolz, A.~Brandenburg and W.~Buchmuller, \emph{{Thermal production of
  gravitinos}},
  \href{https://doi.org/10.1016/S0550-3213(01)00132-8}{\emph{Nucl. Phys. B}
  {\bfseries 606} (2001) 518--544},
  [\href{https://arxiv.org/abs/hep-ph/0012052}{{\ttfamily hep-ph/0012052}}].

\bibitem{Pradler:2006qh}
J.~Pradler and F.~D. Steffen, \emph{{Thermal gravitino production and collider
  tests of leptogenesis}},
  \href{https://doi.org/10.1103/PhysRevD.75.023509}{\emph{Phys. Rev. D}
  {\bfseries 75} (2007) 023509},
  [\href{https://arxiv.org/abs/hep-ph/0608344}{{\ttfamily hep-ph/0608344}}].

\bibitem{Rychkov:2007uq}
V.~S. Rychkov and A.~Strumia, \emph{{Thermal production of gravitinos}},
  \href{https://doi.org/10.1103/PhysRevD.75.075011}{\emph{Phys. Rev. D}
  {\bfseries 75} (2007) 075011},
  [\href{https://arxiv.org/abs/hep-ph/0701104}{{\ttfamily hep-ph/0701104}}].

\bibitem{Eberl:2020fml}
H.~Eberl, I.~D. Gialamas and V.~C. Spanos, \emph{{Gravitino thermal production
  revisited}}, \href{https://doi.org/10.1103/PhysRevD.103.075025}{\emph{Phys.
  Rev. D} {\bfseries 103} (2021) 075025},
  [\href{https://arxiv.org/abs/2010.14621}{{\ttfamily 2010.14621}}].

\bibitem{Kasuya:2001hg}
S.~Kasuya and M.~Kawasaki, \emph{{Q ball formation: Obstacle to Affleck-Dine
  baryogenesis in the gauge mediated SUSY breaking?}},
  \href{https://doi.org/10.1103/PhysRevD.64.123515}{\emph{Phys. Rev. D}
  {\bfseries 64} (2001) 123515},
  [\href{https://arxiv.org/abs/hep-ph/0106119}{{\ttfamily hep-ph/0106119}}].

\end{thebibliography}\endgroup

\end{document}